\documentclass[sigconf]{acmart}

\renewcommand\footnotetextcopyrightpermission[1]{} 
\usepackage[utf8]{inputenc}
\usepackage{subcaption}
\usepackage{multirow}
\usepackage{todonotes}
\usepackage{amsmath,amsfonts}
\usepackage{algorithmic}
\usepackage{enumitem}
\graphicspath{{./figs/}}

\definecolor{ao}{rgb}{0.0, 0.5, 0.0}
\definecolor{amaranth}{rgb}{0.9, 0.17, 0.31}

\AtBeginDocument{%
  \providecommand\BibTeX{{%
    \normalfont B\kern-0.5em{\scshape i\kern-0.25em b}\kern-0.8em\TeX}}}

\copyrightyear{2021}
\acmYear{2021}
\setcopyright{acmlicensed}\acmConference[FAccT '21]{Conference on Fairness, Accountability, and Transparency}{March 3--10, 2021}{Virtual Event, Canada}
\acmBooktitle{Conference on Fairness, Accountability, and Transparency (FAccT '21), March 3--10, 2021, Virtual Event, Canada}
\acmPrice{15.00}
\acmDOI{10.1145/3442188.3445905}
\acmISBN{978-1-4503-8309-7/21/03}

\begin{document}

\title{Designing Accountable Systems}

\author{Severin Kacianka}
\email{severin.kacianka@tum.de}
\affiliation{%
  \institution{Technical University of Munich }
  \streetaddress{Boltzmannstr. 3}
  \city{Garching}
  \country{Germany}
  \postcode{85748}
}

\author{Alexander Pretschner}
\email{alexander.pretschner@tum.de}
\affiliation{%
  \institution{Technical University of Munich }
  \streetaddress{Boltzmannstr. 3}
  \city{Garching}
  \country{Germany}
  \postcode{85748}
}

\begin{abstract}

Accountability is an often called for property of technical systems. It is a
requirement for algorithmic decision systems, autonomous cyber-physical systems,
and for software systems in general. As a concept, accountability goes back to
the early history of Liberalism and is suggested as a tool to limit the use of
power. This long history has also given us many, often slightly differing,
definitions of accountability. The problem that software developers now face is
to understand what accountability means for their systems and how to reflect it
in a system's design. To enable the rigorous study of accountability in a
system, we need models that are suitable for capturing such a varied concept. In
this paper, we present a method to express and compare different definitions of
accountability using Structural Causal Models. We show how these models can be
used to evaluate a system's design  and present a small use case based on an
autonomous car.

\end{abstract}

\maketitle

\section{Introduction}

Accountability differs from many other system properties because it is
notoriously hard to define and its benefits are elusive to name. A property like
performance can often be measured with hard numbers and has the very clear
benefit that more performance means more or faster operations. Security and
privacy are hard to measure, but have the well understood meaning of \emph{keeping
the bad guys out} and \emph{keep data from unauthorized eyes}. In a recent
systematic literature review, \citeauthor{wieringa2020account} has written that
many organizations are ``advocating for more algorithmic accountability,
yet a thorough and systematic definition of the term lacks, and it has not been
systematically embedded within the existing body of work on
accountability''~\cite{wieringa2020account}[p.~2 ]. This finding now begs the question
of what exactly these organizations are advocating if there is not even a
definition of the concept. This feeling of ambiguity is reinforced when looking
at the concrete examples given in the appendix of \cite{wieringa2020account}.
The first is an automatic system that checks if people repay their debt, the
second one is a system that automatically anonymizes permits, the third and
fourth systems check for fraudulent social benefit claims. All examples do
something that some people cannot know or understand and might find
objectionable.  Accountability is now supposed to fix this. Similarly,
\cite{kacianka2017mapping} surveyed implementations of accountability in
computer science and found that systems will often implement \emph{something} and
then just call it \emph{accountability}, without trying to ground that in any
definition or understanding of the term.  In complex systems, accountability is
often deflected and hard to pinpoint, which leads to blame being assigned to
humans \cite{elish2019moral}.  As an example, a ``pilot error'' often is not
just an error of the pilots, but a complex interplay of the humans and the
technical~systems. 

Currently, the literature does not offer any method to model the accountability
of a system, especially across system boundaries. It offers no way to quantify
or qualify the accountability of a system, nor even a precise language to reason
about it or compare implementations.  The current state of the art does not go
beyond giving differing definitions of what accountability might mean and
proposing implementations of accountability in specific contexts. In this paper,
we show how to leverage one commonality among all definitions of accountability,
namely causality, to express accountability definitions and identify them in the
causal model of a system. This allows us to describe patterns in the design of a
system that are necessary to fulfill a specific notion of accountability.
Moreover, this knowledge helps us to reason about what data needs to be logged
and, conversely, what data can be omitted, without compromising the system's
adherence to the chosen definition of accountability.  We propose to use
Structural Causal Models (SCMs) as the mathematical foundation.  They are
flexible enough to model even the most complex systems and offer a toolbox of
mathematical methods to analyze them.

\section{Background}
\label{sec:whatis}
\label{sec:many_acc}
Accountability is a concept rooted in Liberalism and was first introduced by
political philosophers like John Locke and Adam Smith, who used it in the
17\textsuperscript{th} and 18\textsuperscript{th} century to describe the fact
that official representatives will have to justify their action to someone,
ultimately their sovereign.\footnote{See \cite{dubnick2010moral},
\cite{bovens2007analysing}, or  \cite{lindberg2013mapping} for a more detailed
history.} This core idea was then  picked up and refined by other political and,
later, social scientists.  In a survey, Lindberg 
gives the central idea as ``when decision-making power is transferred from
a principal (e.g.  the citizens) to an agent (e.g. government), there must be a
mechanism in place for holding the agent accountable for their decisions and
tools for sanction''~\cite[p.~203]{lindberg2013mapping}.
  Bovens writes that
``[t]he most concise description of accountability would be: ‘the
obligation to explain and justify conduct’, '' while also warning that
``[a]s a concept, however, ‘accountability’ is rather elusive. It has
become a hurrah-word, like ‘learning’, ‘responsibility’, or ‘solidarity’, to
which no one can object''~\cite[p.~7]{bovens2007analysing}.

This core idea is, with some variations, deeply embedded into the fabric of
liberal democracies. From the idea that voters will hold politicians accountable
for their performance, to companies that are accountable to their shareholders,
to the legal system, where wrongdoing is discouraged by the possibility of
being held accountable for one's actions. As such, accountability rose to
prominence in computer science together with the tight integration of computers
into our societies and their increasing effect on daily life, for example by
managing medical records or controlling vehicles. 

This long history of the term has led to many different definitions and meanings
of accountability. Lindberg, for example identified twelve different subtypes of
accountability and has also cautioned us that ``[i]t cannot be assumed that
findings in the area of one subtype of accountability are relevant for
another''~\cite[p.~204]{lindberg2013mapping}. For example, if we find an
implementation of accountability that works well in a societal setting, it is
not a given that it will also work in a legal setting.
\cite{kacianka2017mapping} have found a tendency in computer science to not
worry much about the underlying definitions of accountability. Even recent
works, for example \cite{wieringa2020account}, usually pick some definitions and
declare it as typical. This is, in our opinion, wrong.  Computer science
should not try to pick winners, and push one theory over another.  When talking
about accountability, it is important to be precise about the exact meaning. As
Lindberg puts it, ``everything is not accountability: it is but one of many
possible ways to constrain the (mis-)use of
power''~\cite[p.~202]{lindberg2013mapping}. Other means of limiting power are
the ``devolution of power, violence, economic pressure, public shaming, and
anarchy''~\cite[p.~205]{lindberg2013mapping}. However, since accountability is a
very old concept, it has multiple meanings that often have subtle differences.
This is why, when talking about \emph{accountability} or \emph{making systems
accountable}, we should always first try to define what we actually mean. For
example, in some definitions sanctioning an actor for their action is considered
part of accountability, while in others a principal can only sanction an agent
if they do not provide an account. Such differences have a huge impact on the
underlying system design and should thus be made explicit, and not left
ambiguous by just using the term \emph{accountability}. 
In our view, all definitions of accountability have some merit in a specific
context, and computer science should strive to offer ways to implement any
definition. It is the purview of fields like sociology or the political sciences
to debate the intricacies of the definitions themselves.\footnote{Causes and
accountability are also an important topic in law. Since causality has only been
formalized very recently, relevant literature can be found under the term
statistics, e.g., \cite{degroot1986statistics}.} They have accumulated
experience in debating these finer points and computer science should rely on
their insights and offer ways to realize the theories developed there. Here we
will now present some approaches to accountability as an example of their wide
variety and show how to formalize them later in Section~\ref{sec:formalize}.

\subsection{Lindberg}
\label{sec:lindberg}

Staffen Lindberg \cite{lindberg2013mapping} surveyed the literature in the
social sciences and distilled the following definition of accountability:  
\begin{itshape}
\begin{enumerate}
\item An agent or institution who is to give an account (A for agent);
\item An area, responsibilities, or domain subject to accountability (D for
domain);
\item An agent or institution to whom A is to give account (P for principal);
\item The right of P to require A to inform and explain/justify decisions with 
regard to D; and 
\item The right of P to sanction A if A fails to inform and/or explain/justify 
decisions with regard to D~\cite[p.~209]{lindberg2013mapping}.
\end{enumerate}
\end{itshape}

The first two points mean that there is an agent that has some power in a
certain domain and knows that they need to give an account for their actions.
The third and the fourth condition imply that there is a third party that has
the right to require $A$ to explain and justify their decisions. The last
condition requires that $P$ can sanction $A$. Lindberg adds an important
restriction, often lost in other definitions: he distinguishes between the right
of $P$ to sanction $A$ for not providing the information and the right to
sanction $A$ for the content of effect of an action. Another important
implication is that there needs to be ``standard or measurable
expectations''~\cite[p.~211]{lindberg2013mapping}   to have accountability.
Without a clear idea of what is acceptable and unacceptable behavior, it cannot
be evaluated and sanctioned. As such, we always need some form of evidence.

\subsection{Bovens}
\label{sec:bovens}
Mark Boven's definition \cite{bovens2007analysing} became popular recently in
computer science because it was used as the definition in the systematic
literature review conducted by \cite{wieringa2020account}.  Bovens finds that
accountability is hardly defined and he tries to counteract this vagueness and
make it ``more amendable to empirical
analysis''~\cite[p.~7]{bovens2007analysing}   He focuses on \emph{public
accountability} and gives a short definition, ``the obligation to explain and
justify conduct''~\cite[p.~9]{bovens2007analysing}, before giving the more
detailed one, as follows:

\begin{itshape}
\begin{enumerate}

\item There is a relationship between an actor and a forum
\item in which the actor is obliged
\item to explain and justify
\item his conduct,
\item the forum can pose questions,
\item pass judgement,
\item and the actor may face consequences~\cite[p.~12]{bovens2007analysing}.

\end{enumerate}
\end{itshape}

Following his definition, actors can be individuals or organizations,  and a
forum can also be a specific person, an organization, or even the general
public. The relationship between actor and forum will often, but not always, be
a principal-agent relation in which the forum delegates power to the agent, who
is then held to account. The obligation of the actor might be formal or
informal.  The act of giving an account consists of three stages. First, ``the
actor is obliged to inform the forum about his conduct, by providing various
sorts of data about the performance of tasks, about outcomes, or about
procedures''~\cite[p.~10]{bovens2007analysing}. Second, `` there needs to be a
possibility for the forum to interrogate the actor and to question the adequacy
of the information or the legitimacy of the
conduct''~\cite[p.~10]{bovens2007analysing}. Finally, ``the forum may pass
judgement on the conduct of the actor''~\cite[p.~10]{bovens2007analysing}.
Additionally, he also requires the possibility of consequences for the actor if
they do not comply with the requests of the forum.  One fundamental difference
between the definitions of Lindberg and Bovens is that Bovens requires the actor
to regularly update the forum, whereas Lindberg suggest that the principal can
demand an account from the agent at any time.

\subsection{Hall}
\label{sec:hall}
Hall et al. \cite{hall2017psy_acc} look at accountability from the perspective
of psychology. As such they focus on what it means for an individual to
\emph{feel accountable}. Their exact field of study is \emph{felt
accountability}.  In their overview, they find that at the core of
accountability is the expectation that one's actions will be evaluated. They
emphasize that it is not necessary that this evaluation does occur, but that the
possibility that an evaluation occurs must be present. Furthermore, the actor
needs to believe that an account-giving (i.e., an explanation) might be
required. This account is then given to a salient audience that might reward or
sanction the agent's behavior. 

In their review of models of accountability, they find that accounts are often
used by agents to protect their self-image and develop their social identity.
This underscores the important role accountability has in a society and supports
the assumption that complex societies and social order necessarily need
accountability to augment the reduced level of personal trust between
individuals. In their survey, they describe four essential features of
accountability. First, the \emph{accountability source} describes to whom one
feels accountable. Second, the \emph{accountability focus} captures how things
get done and how they relate to the results. Third, \emph{accountability
salience} expresses how important the task is for which an agent might be held
accountable; the idea is that agents will be more careful if their action is
more significant. Fourth, and finally, \emph{accountability intensity} captures
for how many things an agent is accountable; here it is thought that being
responsible for multiple things increases stress.


\subsection{RACI}
\label{sec:raci}

The organizational sciences have developed several practical
frameworks to understand accountability in an organization. Here, we present the
Responsible-Accountable-Consult-Inform (RACI) framework
\cite{racibook,smith2005role}. Such frameworks, sometimes called tools, are used
in practical settings to explicate accountability relationships in teams and
organizations. While these tools do not build on a sophisticated body of
scientific literature, they are nonetheless important in practical settings
because they allow people to express their understanding and perception of
accountability in a given setting. We assume that such practical approaches will
often be the basis for accountability expectations of systems, and thus need to
be considered in any attempt to formalize accountability for them. 

RACI, specifically, tries to explicate the roles of people in an organization
and helps to reconcile the conception of a role, i.e., what a person thinks they
are doing, with the expectation of a role, i.e., what others think the person is
doing, and with the behavior of a role, i.e., what the person actually does.
Following \cite{smith2005role}, having a RACI matrix helps to align these three
aspects. However, they also point out that this is an ongoing process that needs
to constantly realign those three aspects, whenever they drift apart. They list
a few typical signs, such as ``Questions over who does what'' or ``Concern over
who makes decisions''~\cite[p.~4]{smith2005role}, that arise regularly during
the design of systems.  Among other things, one major difference of this
definition from the others is the aspect of consultation, once again underlining
our finding in the introduction that definitions are manifold and different.
\cite{smith2005role} define the following four aspects in the RACI framework:

\begin{itshape}
\begin{itemize}
\item \textbf{R}esponsible: The individual who completes a task. Responsibility
can be shared.
\item \textbf{A}ccountable: The person who answers for an action or decision.
There can be only one such person.
\item \textbf{C}onsult: Persons who are consulted prior to a decision. 
Communication must be bidirectional. 
\item \textbf{I}nform: Persons who are informed after a decision or action is
taken. This is unidirectional communication.  

\end{itemize}
\end{itshape}

\subsection{Computer Science}
\label{sec:acc_cs}
A landmark publication on accountability in computer science was published by
Weitzner et al.~\cite{Weitzner:2008}.\footnote{The core problem, however, was
discussed much earlier. Notable contributions technical solutions are Lamport's
logical time stamps in 1978 \cite{lamport1978time} while Nissenbaum's discussed
the ``eroding accountability in computerized societies'' in 1996
\cite[p.~25]{nissenbaum1996accountability}. } They provided a definition for
\emph{Information Accountability}, as an improvement on classic preventive data
control measures.  Classically, systems ensure a user's privacy by ensuring that
data cannot be accessed by unauthorized personnel and thus prevent data leaks.
Weitzner et al.  changed this premise and, drawing parallels to law enforcement,
suggested to build systems in such a way that it is easy to trace data leaks and
then leverage the existing legal system to punish misbehavior.  This idea was
later refined and formalized by Feigenbaum et al.
\cite{feigenbaum2011towards,feigenbaum2012systematizing}, albeit with a focus on
security and not privacy. Coinciding with the discussion on e-voting systems,
K{\"u}sters et al.~\cite{kusters2010accountability} formalized accountability in
relation to verifiability. Here, the main question is how to design an e-voting
system such that the results can be trusted and any attempts to falsify the vote
count or the votes will be detected and the perpetrator held to account.  The
A4cloud project \cite{a4cloud2014}, coinciding with the spread of cloud services
into society and questions about data protection, has done extensive work on
accountability in cloud environments, with a focus on data protection and
privacy. They offer a reference architecture, tools to complete certifications,
and risk assessment.  Looking at their
website,\footnote{\url{http://a4cloud.eu/Accountability.html}} they defer the
exact definition of accountability to contracts or service level agreements.

Furthermore, Kacianka et al. \cite{kacianka2017mapping} conducted a systematic
mapping study to understand how accountability is understood and implemented in
research tools. In this study, they identified a steady rise in publications on
the subject and found that most research was either a solution proposal or an
evaluation of an approach. Analyzing the prominent application domains, they
found that cloud computing was clearly dominant, followed by distributed data
sharing and web applications. The most prominent use cases were privacy focused,
in line with \cite{Weitzner:2008}, and the most popular techniques were
cryptographic and network protocols, with some dedicated accountability
protocols as well. Obviously, the focus on information accountability makes
related definitions different from those discussed before.

\subsection{Algorithmic Accountability}
The term algorithmic accountability first gained prominence with the paper by
Nicholas Diakopoulos~\cite{diakopoulos2015algorithmic} where he discussed how
journalists might investigate algorithms that started to make more decisions
that affected human lives. In this vein, the literature on the subject usually
focused on the understanding of machine learning algorithms. Examples include
algorithms used in  court decisions~\cite{angwin2016machine},
policing~\cite{kaufmann2018predictive}, and similar settings where human lives
are directly affected by opaque computer systems. Most of the literature is
highly critical of these systems, using terms like  \emph{Weapons of Math
Destruction}~\cite{o2016weapons} or \emph{Algorithms of
Oppression}~\cite{noble2018algorithms}.  The general approach to counter the
power of algorithms is to make the decisions of algorithms transparent and
explainable. Recently, Maranke Wieringa surveyed the literature on algorithmic
accountability and found that  ``[w]hat is denoted with algorithmic
accountability is this kind of accountability relationship where the topic of
explanation and/or justification is an algorithmic
system''~\cite[p.~2]{wieringa2020account}. Her survey also finds that typically
algorithmic accountability follows the definition of Mark Bovens
\cite{bovens2007analysing} given above. 

In earlier works, transparency was often seen as a solution, but Ananny and
Crawford~\cite{ananny2018seeing} have shown that transparency alone is not
sufficient for accountability. Amongst other reasons, a main point is that we
also need someone to understand the output of such a transparency mechanism. To
alleviate this problem,  Wachter et al.~\cite{wachter2017counterfactual}, and
later Tim Miller~\cite{miller2018explanation} as well as Mittelstadt et
al.~\cite{mittelstadt2019}, have proposed using contrastive explanations to make
decisions understandable for humans.  Miller gives the example of a machine
learning classifier that categorized an insect as a spider or a beetle. An
explanation it would give is that a result is categorizes as a spider because it
has eight legs instead of six. In contrast to the weights in a neural net or the
layout of a decision tree, such an explanation would be useful and
understandable for a human. 

\section{Causality}

The study of causality\footnote{We closely follow the definition of SCMs as
introduced and refined over the years by Judea Pearl:
\cite{pearl2000causality,pearl2009causality,pearl2016causal,pearl2018book} and
Joseph Halpern \cite{halpern2016actual}. Other definitions of causality exist
and might sometimes even be more suitable, but Pearl's SCM approach is the most
widely adopted notion. Even more, \cite{halpern2016actual} discusses several
incompatible notions  of actual accountability and there are arguments that a
single definition is not useful  \cite{glymour2010actual}. } is of a specific
interest to the study of accountability, our main subject matter, because
causality is a prerequisite for accountability.  In understanding how causal
effects work, we can improve the design of our systems to make sure that effects
of causes are clearly understood and then in turn ensure that the causes of
effects are easy to identify. The first notion is prospective and the second one
retrospective.\footnote{See also the notes in the first chapter of
\cite{halpern2016actual}.} Both are deeply intertwined, but often only studied
separately. In this paper, we combine the study of both and build on the
assumption that a good prospective causal model is also a good retrospective
causal model. For prospective models, we can find structures and patterns that
allow us to show that some variables are not relevant to certain outcomes and
once a specific outcome comes to pass, we can use retrospective reasoning to
identify the concrete cause, relative to the given context.  

Historically, causality and causal relationships are tightly connected to
statistics.  However, as laid out by \cite{pearl2000causality}, whereas
statistical relations are epistemic and describe what we know or believe about
the world, causal relationships are ontological, meaning that they describe
objective constraints on the world. This means that causal relationships are
much more stable and should not change if the environment changes. Still,
causality is tightly linked to statistics as many ``causal statements are
uncertain'' \cite{pearl2016causal} and are thus often only true with some
probability.\footnote{For example, ``smoking causes cancer'' is true, but a
single cigarette is very unlikely to cause cancer.} As such, many prospective
causal models will answer questions with a certain probability. For
accountability, we often need exact and retrospective answers.  The question
\emph{Did Alice cause the crash?} should have a clear retrospective answer. For
this we use Actual Causality \cite{halpern2016actual}. It allows us to use a
(prospective) causal model and reason about it in a specific context.  For
example, we might have a prospective model that shows that texting while driving
causes accidents (with a certain probability). Then we might have the specific
context of an accident in which we know that Alice was distracted because she
was texting. We can then set the context for this accident and use actual
causality reasoning to find the cause of the accident; in this case Alice caused
the accident by being distracted. 
In the literature, \cite{datta2015program} as well
\cite{chockler2004responsibility} already suggested using actual causality as a
building block  for accountability, and \cite{smc2018} show how causality can be
useful to model accountability. 

\subsection{Type Causality}

\begin{figure}
	\centering
	\begin{subfigure}[t]{.5\textwidth}
		\centering
		\includegraphics[height=0.75cm]{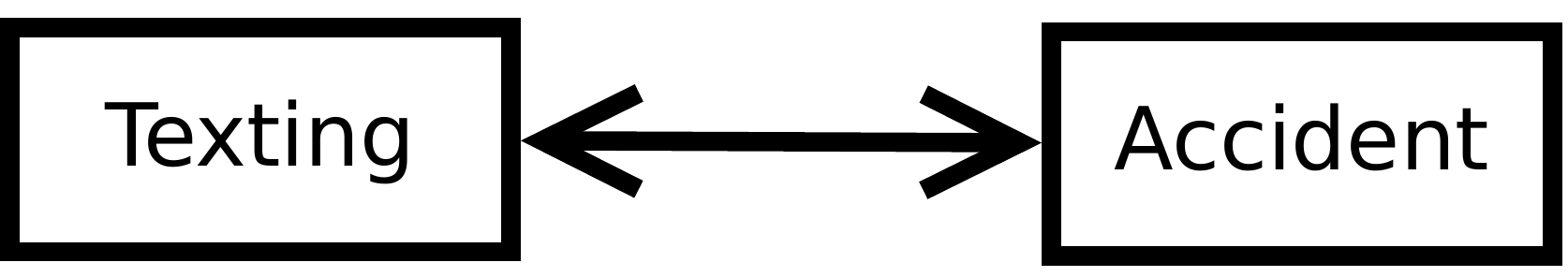}
		\caption{A correlation between texting and accidents.
		}\label{fig:texting1}		
	\end{subfigure}
	\quad
	\begin{subfigure}[t]{.5\textwidth}
		\centering
		\includegraphics[height=0.75cm]{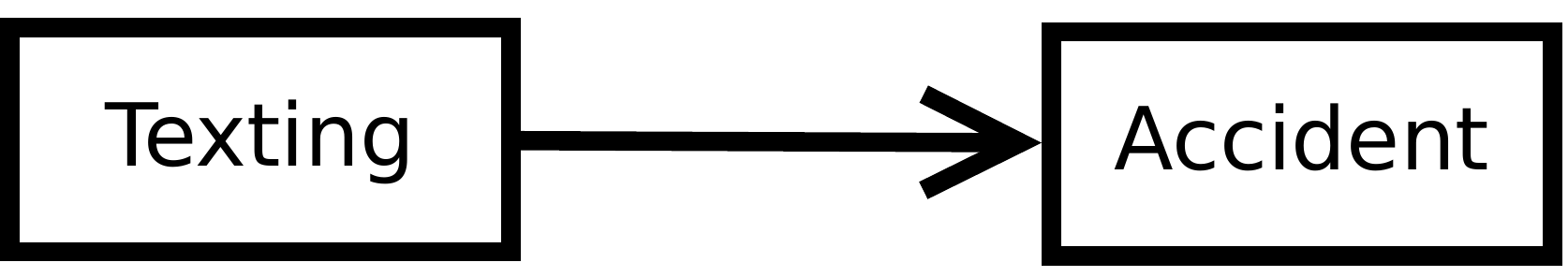}
		\caption{Our model assumes that texting is the
		cause.}\label{fig:texting2}
	\end{subfigure} 
	\quad
	\begin{subfigure}[t]{.5\textwidth}
		\centering
		\includegraphics[height=0.75cm]{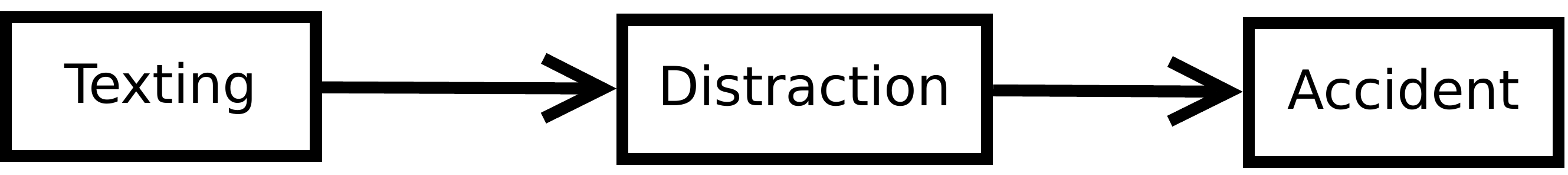}
		\caption{We discover the mechanism by which texting causes accidents.
		}\label{fig:texting3}
	\end{subfigure} 
	\caption{From correlations to causal models. }	
	\label{fig:texting}
\end{figure}

Investigations of causality usually start with the identification of a
correlation between two variables. In Figure~\ref{fig:texting1}, for example, we
notice a correlation between texting while driving and the number of accidents.
A correlation has no direction, and many correlations will turn out to be
spurious, so caused by some unknown third variable, often called a confounder.
The goal of scientific investigations is now to find which correlations are the
result of genuine causal mechanisms in the real world. In the end, such an
investigation will yield a causal model. In our example,
Figure~\ref{fig:texting2} expresses the understanding that texting while driving
causes accidents. Maybe just with a certain probability and under certain
assumptions, but we clearly state that one is the cause of the other. This is
the modeler's understanding of a mechanism in the world. The advantage of
stating it as a causal model is that all assumptions must be made explicit and
that it can be tested against actual data. It might also be refined over time;
Figure~\ref{fig:texting3} shows a causal model that assumes that texting does
not directly cause accidents, but that it does so via a mediator, namely
distraction. Such details are often very important, as they improve our
understanding of the problem and allow us to develop ways to affect, and often
prevent, specific outcomes by targeting the mediators directly.

Such causal relations can be formalized in so-called Structural Causal
Models (SCMs)\footnote{Causal models are also often specified over probability
distributions. For simplicity, we stick to the discrete definition in the paper
and the examples. }
\cite{pearl2016causal}. They are are derived from structural equation
models (SEMs) (e.g., \cite{lomax2004beginner}), but their relations have a
direction. Following Pearl \cite{pearl2016causal}, an SCM consists of two
sets of variables, $\mathcal{U}$  and $\mathcal{V}$ and a set of functions,
$\mathcal{F}$, that assigns each variable in $\mathcal{V}$ a value based on the
value of the other variables in the model. Formally,

\begin{definition}[Structural Causal Model]

\label{def:scm}
	A structural causal model $M$ is a tuple $M = \mathcal{(U,V,F)}$, where
	\begin{itemize}
	    \item $\mathcal{U}$ is a set of \emph{exogenous} variables,
		\item $\mathcal{V}$ is a set of \emph{endogenous} variables, 
		\item $\mathcal{F}$ associates with each variable $X \in
		\mathcal{V}$ a function that determines the value of $X$  given the 
		values of all other variables.
	\end{itemize}

\end{definition}

Every SCM is associated with a \emph{graphical causal model} called the
\emph{graphical model} or the \emph{graph}. Nodes are the variables; edges represent a
causal relationship between them. While the graph does not include the details
of $\mathcal{F}$, its structure alone is enough to identify patterns and causes.
In an SCM, exogenous variables, denoted by $\mathcal{U}$, are external to the
model, meaning that we chose not to explain how they are caused.  They are the
root nodes of the \emph{causal graph} and are not a descendant of any other
variable.  Endogenous variables, denoted by $\mathcal{V}$, are descendants of at
least one exogenous variable and model components of our system and the world
for which we want to explain causes.  $\mathcal{F}$ describes the relationships
between all those variables.  If we knew the value of every exogenous variable,
we could use $\mathcal{F}$ to determine the value of every endogenous variable.
In a graphical model every node represents an endogenous variable and arrows
represent functions from $\mathcal{F}$ between those variables.

\subsection{Actual Causality}
\label{sec:actual}

Type causal models make predictions on how events will unfold. This is useful
because we want to build systems in a way that they \emph{will probably be}
accountable. To achieve this, we try to make sure that the causal effects within
a system are clearly understood and that the structure ensures that as many
components as possible are causally independent. However, what if an unwanted
event has already happened? In this case we do not care about the probabilities
of events; we know they happened, but we want to find out why exactly they did
happen. For this we need the concept of \emph{actual causality}
\cite{halpern2016actual}. It is backward-looking and stands in contrast to
\emph{type causality} which is forward-looking.\footnote{Note that actual
causality can also deal with probabilities for retrospective events; see
\cite[Ch. 2.5]{halpern2016actual}.  }

\begin{figure}[h!]
	\centering
	\includegraphics[width=0.25\textwidth]{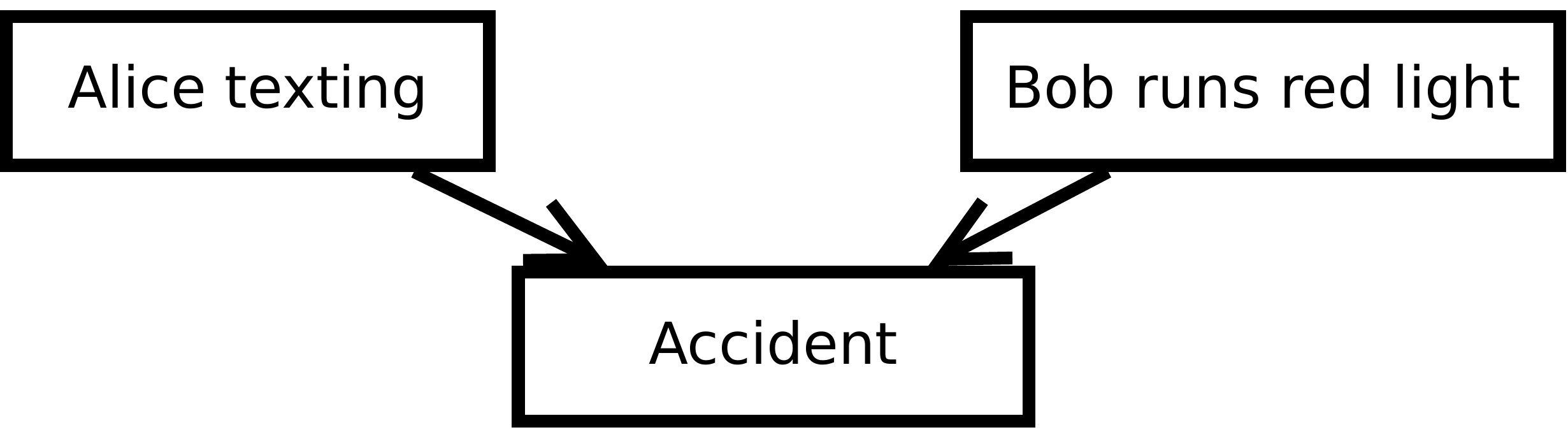}
	\caption{Two cars causing an accident.  }
	\label{fig:suzy_billy1}
\end{figure}

To illustrate actual causality, Figure~\ref{fig:suzy_billy1} depicts an accident
between two cars.\footnote{This is based on the classic Suzy-Billy rock throwing
example \cite[Example 2.3.3]{halpern2016actual}.} Imagine two drivers, Alice and
Bob, breaking the law at an intersection. Alice is texting and thus distracted
while Bob accidentally runs over a red light.  This example is designed to show
that the simple but-for test\footnote{The but-for test is a simple understanding
of causality that reads ``$A$ is a cause of $B$ if, but for $A$, $B$ would
not have happened'' \cite{halpern2016actual}. It is often used in the legal
context, called by its Latin name \emph{sine qua non} test. In this domain,
several improvements were developed such as the INUS (an Insufficient but
Necessary element of an Unnecessary but Sufficient set) or the NESS (Necessary
Element of a Sufficient Set) test.  For a detailed overview, see
\cite{sinequanon}.}  is not adequate to attribute causality. Here, intuitively
both Alice and Bob are necessary for the accident to happen. However, had
Alice not been texting, Bob would still have run the red light and caused the
accident. So the accident would have happened, no matter what Alice did,
suggesting her behavior is not a cause, which goes against our understanding of
causality.  The goal of the Halpern-Pearl definition (see Appendix~\ref{sec:hp}
for its formalization) of causality is  to find a precise mathematical
definition to enable algorithmic reasoning over such examples so that the
result conforms with our human understanding of causality. 

\subsection{Causality and Accountability}
It is now important to note that SCMs can describe purely technical systems.
They do not require a principal or any human at all. For a system to be
accountable, however, we require a natural or legal person that is not just a
cause for an effect, but is accountable for that effect. In other words,
causality is necessary for accountability, but by itself it is not sufficient
for accountability. The additional requirements are given by accountability
definitions such as the ones introduced in Section~\ref{sec:many_acc}. 
In our work with SCMs,
we found the following differentiation of terms useful:

A \textbf{cause} is the \emph{actual cause} in an SCM as determined by the
Halpern-Pearl definition of actual causality. Similarly, \textbf{to cause}
means that an endogenous variable in an SCM is the \emph{actual cause} of
another endogenous variable. Causes are purely technical, without any notion of
intent or other social attributions.\footnote{However, it is important to note
that causes are always relative to a causal model. The causal model might be
biased and thus social attributions can leak into the model.}

\textbf{Responsibility} is the commitment of an entity to act a certain way and
the ability to affect or change an outcome.\footnote{When this commitment is
derived from moral reasons, the term ``duty'' will often be used.} This entity
is \textbf{responsible} for a certain outcome. As such we have explicit notions
of \emph{normality} that are used to determine responsibility. This entity is
then responsible for an outcome when, had it acted \emph{normally}, the outcome
would not have happened.    

\textbf{Blame} is a social process in which an agent has a specific notion of
normality and will find fault with some entity for the fact that this normality
is violated. This agent \textbf{blames} that entity for some outcome, even if
that entity is not aware of this notion of normality and thus might not be
responsible for the outcome. 

A \textbf{transparent} system will have an SCM available and log enough data to
set the context of the causal model after some event. \textbf{Transparency}
indicates that an SCM is available, although it makes no statement about the
quality of the SCM.  

\textbf{Accountability}, finally, means that we have a natural or legal person,
called an agent, that is \emph{responsible} for some outcome. This
\emph{responsibility} is made \emph{transparent} with an SCM, and thus allows a
dedicated principal to ask this agent for his, her, or their account and
\emph{blame} this agent for an unwanted outcome. So we can actively question the
agent and understand him, her, or them. This means that our attribution of blame
is no longer purely subjective, but derived from objective facts.  An
\textbf{accountable system} is a socio-technical system in which the
\emph{responsibility} for every outcome is linked to an agent, so a natural or
legal person. It has an \textbf{accountability mechanism} which is an extension
of the system that helps the principal to keep the agent accountable. Lastly,
an \textbf{accountability definition} describes the necessary structures in the
SCM.

We do not claim that our definitions fit all situations and it is not our
intention to obfuscate the long history of these terms. Yet, we believe that
it is helpful to clearly state our understanding so that it can be compared,
discussed,  and contrasted to others.

\section{Accountability Structures in Causal Models}
\label{sec:formalize}
 
With SCMs as the means to formalize causal models, we can now revisit the
accountability definitions given  in Section~\ref{sec:many_acc} and look at them
through a \emph{causal lens}. Unfortunately there is no automatic, deterministic
way of translating them into SCMs. Causal models express the modeler's
understanding of the subject matter and their advantage is that they are
unambiguous. Here, we do not argue that our translations to SCMs
are perfect. Our point is that they are easy to understand, precise in their
meaning, and thus enable a discussion and review of a given definition of
accountability. 

SCMs are useful because they may exhibit  patterns, such as chains (see
Appendix~\ref{sec:chains}), and specific structures, such as the Front- and
Backdoor Criterion (see Appendix~\ref{sec:analyzing}) that allow us to show that
some nodes will have no causal influence on a specific event. To leverage this,
first, the models need to contain actions taken by humans, and as our purpose
is the design of accountable systems, also actions taken by machines.  This
requires us to express the accountability relation as a set of variables that
causally influence each other. On the level of accountability definitions, we do
not care about the exact nature of this influence, so we do not need to specify
$\mathcal{F}$. The reason for this is that if there is a causal influence,
accountability might be necessary and our system should provide data to ensure
it.  It is only after something unwanted has happened that we need
$\mathcal{F}$ to understand the cause and answer questions of accountability.
Conversely, if we can show that a specific variable, representing the actions of
an agent, cannot contribute to a specific outcome, this agent cannot be
accountable for that outcome.

\subsection{Lindberg, Bovens, and Hall}
\label{sec:lindberg_formal}
\begin{figure}
	\centering
	\includegraphics[width=0.35\textwidth]{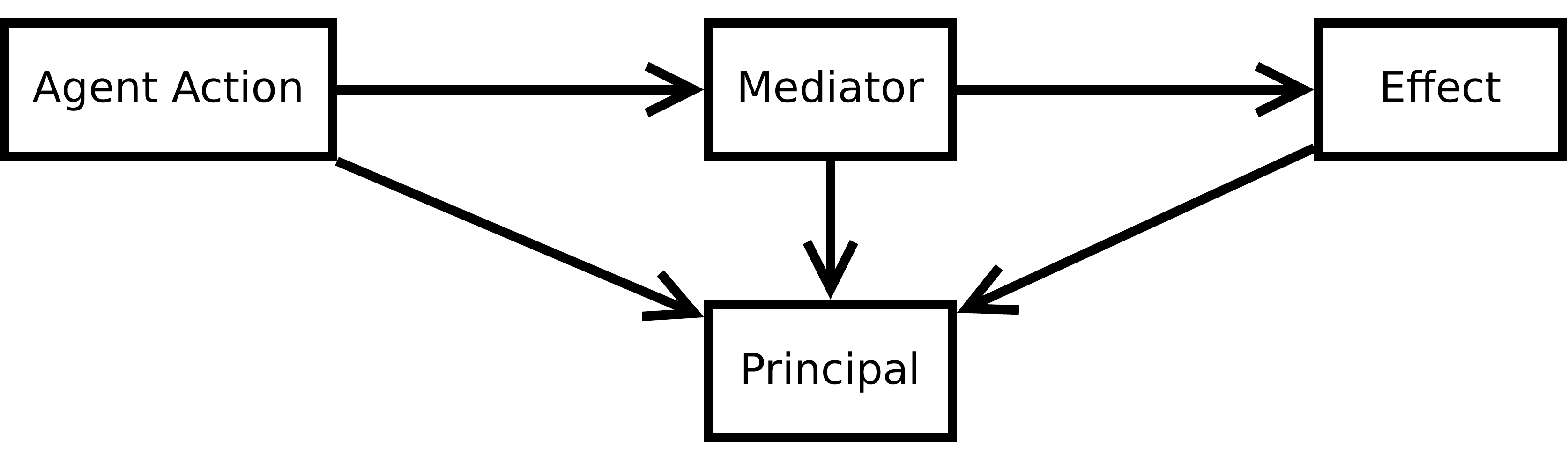}
	\caption{The causal model for Lindberg's definition.}
	\label{fig:lindberg_cm}
\end{figure}

Lindberg's definition of accountability requires an agent, $A$, that should give
an account for some effect, $E$, caused by $A$. Translated to an SCM, this means
we need to  have at least the relation $A \rightarrow E$ in the model. $A$ is a
representation of the action taken by the agent and the result of that action,
$E$, will depend on some value $A$ takes.  To allow for the fact that effects
are often not caused directly, but indirectly via a mediator ($M$), we would
make the possible use of a mediator explicit by adding the relation $A
\rightarrow M \rightarrow~E$.  An example for a mediator is a power steering
wheel that amplifies and translates the movements of the driver ($A$) into the
actual angle of the wheels ($E$).  Next, Lindberg requires a domain that is
subject to accountability. This, conveniently, is captured very precisely by the
model itself. It reflects the context in which $A$ is embedded, as well as the
effect $A$ might cause. Finally, the definition contains a principal, $P$, that
transferred power to $A$ and thus has the right to demand information from $A$
and, should $A$ not comply, sanction~$A$. Figure~\ref{fig:lindberg_cm} now
depicts the structure of the Lindberg pattern. $P$ will be causally affected by
$A$ and also might be affected by $M$ and $E$. In Lindberg's view, the principal
is not directly involved in the course of events. Moreover, typical actions by a
principal, such as helping in the design of the system or investigating an
accident, are beyond the scope of the technical system and part of the society
the system is embedded in. To reflect this, the models shows no arrows
originating from $P$. Any action taken by $P$ goes beyond the limits of the
technical system. In other words, it is necessary for a system to exhibit the
pattern in Figure~\ref{fig:lindberg_cm}, but it requires additional facilities
in the social world around the system to be accountable.

One interesting property of this pattern is that it seems to be at the core of
several other definitions. Despite the differences in the details such as the
timing, Bovens (see Section~\ref{sec:bovens}) shares the same causal model.
Where Lindberg calls for an \emph{agent} giving an account to a
\emph{principal}, Bovens considers an \emph{actor} that \emph{explains his
conduct} to a \emph{forum}. Both the \emph{principal} and the \emph{forum} might
sanction or judge the \emph{agent} or \emph{actor}. This similarity of
definitions suggests to us that the causal models should also be similar. One
pronounced difference is that Bovens requires the actor to regularly inform the
principal about changes, whereas Lindberg sees the principal as asking for
information. In contrast to both, at the core of Hall's definition (see
Section~\ref{sec:hall}) is an agent's expectation that their action will
potentially be evaluated by a third party. As such it also suggests the $A
\rightarrow P$ relation, but with the added twist that $A$ only needs to
\emph{believe} this relation to exist. It does not matter if it exists in
reality. Here, the technical system does not have to provide any logging or
data, so long as $A$ does not know this. Examples are systems that promise to
randomly audit certain transactions, such as tax agencies or anti-cheat tools in
online games. While in a concrete instance there might be no technical means to
evaluate $A$, the system will deter $A$ from misbehaving by introducing the fear
of an evaluation.

\subsection{RACI}
\label{sec:raci_formal}

\begin{figure}
	\centering
	\includegraphics[width=0.45\textwidth]{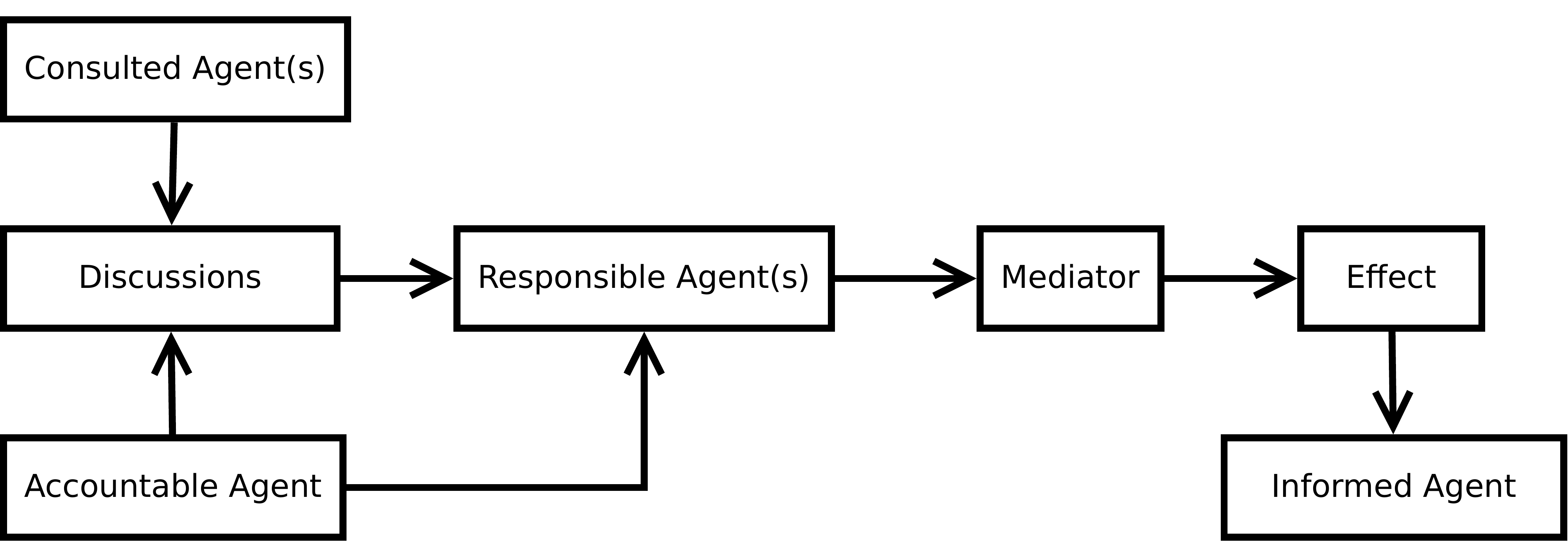}
	\caption{The RACI accountability pattern.  }
	\label{fig:raci_cm}
\end{figure}

RACI (see Section~\ref{sec:raci}) in contrast does not so much look at the
individual, but at an organization as a whole. Similar to Lindberg above, it
features agents that cause some effect, and thus also exhibits the familiar
causal chain $A \rightarrow M \rightarrow E$. It specifically extends the
pattern with an accountable agent, $AA$, that instructs $A$ to do a specific
task. To reflect that in the model, we need to extend it with an edge $AA
\rightarrow A$.  Furthermore, RACI requires any consulted agents, $C$, to be
reflected in the model. We would model this by adding a dedicated node, $D$, to
the SCM to capture the outcomes of these discussions. Lastly, RACI considers
dedicated agents, $I$, that are informed of the effects. In contrast to the
others, RACI requires no dedicated principal.  Figure~\ref{fig:raci_cm} shows
the complete model.

\subsection{Designing an Accountable System}
The question now is, how can we use these accountability structures in the
development of a system? Here, we assume that we have an accurate SCM of the
socio-technical system available. We will   call this $\mathcal{M}$ connoting
that it is a model of the system.\footnote{Getting the SCM is not easy; however,
parts of it can be automated by using models of a system such as fault
and attack trees \cite{nfm} or models of human behavior \cite{crest}.}
$\mathcal{M}$ can be used to answer questions of causality of the system. This,
however, is not the same as answering questions of accountability.  For a system
to be accountable, it needs to conform to a given definition of accountability,
which we call $\mathcal{D}$, connoting that it is a definition. For example, if
our notion of accountability is \emph{the driver is always accountable for what the
car does}, a causal model that does not contain that the driver violates this
notion of accountability.

\begin{figure}
	\centering
	\begin{subfigure}[t]{0.4\textwidth}
		\centering
		\includegraphics[height=.5cm]{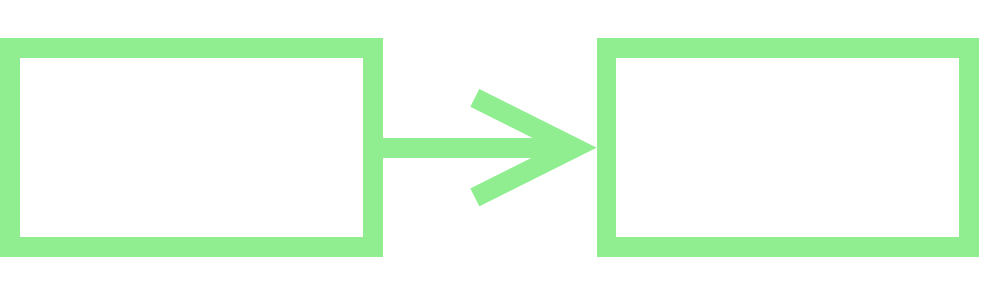}
		\caption{A possible model of an accountability definition. }
		\label{fig:structure2}		
	\end{subfigure}
	\quad
	\begin{subfigure}[t]{0.4\textwidth}
		\centering
		\includegraphics[height=2cm]{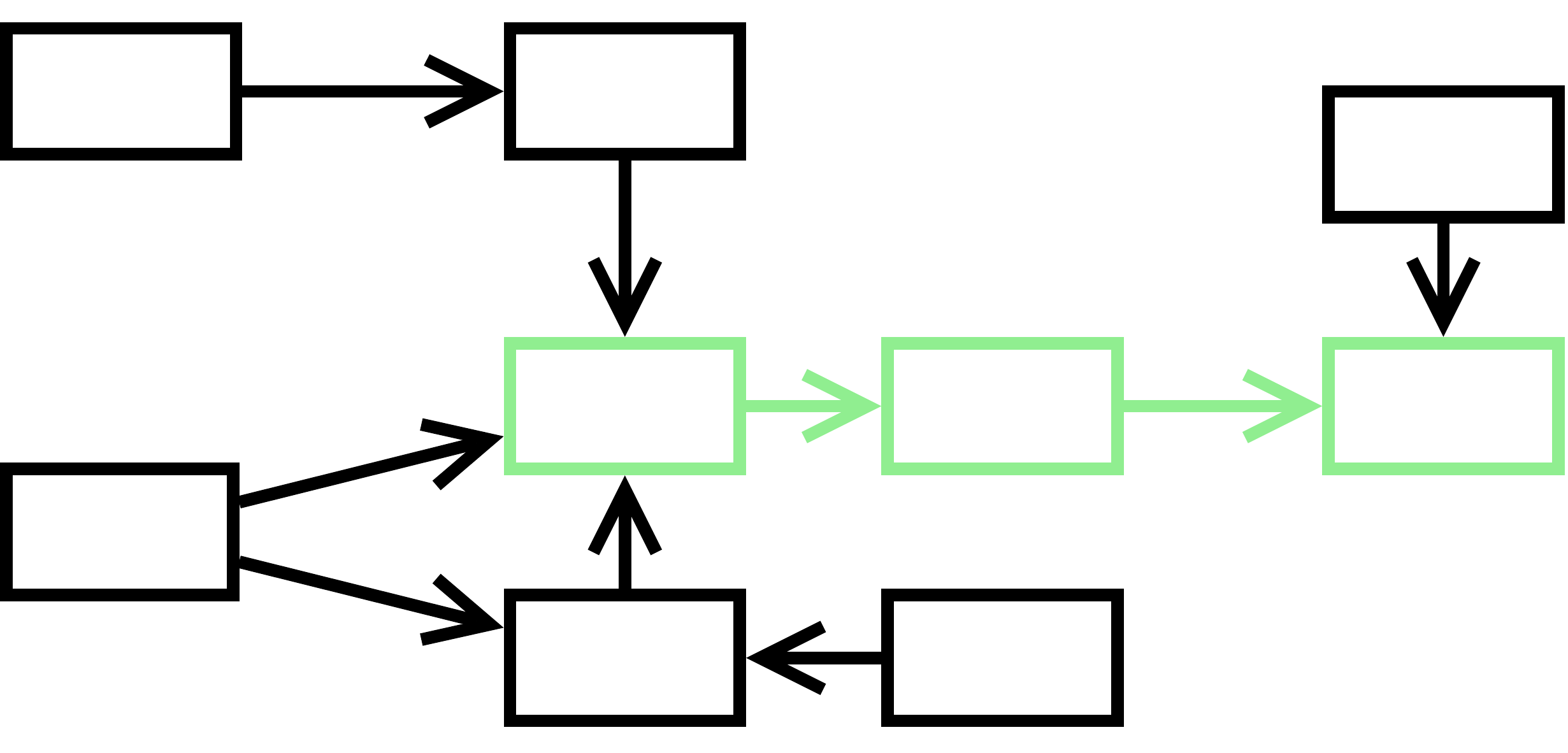}
		\caption{A system model with the equivalent 
		pattern highlighted in green.}
		\label{fig:structure1}
	\end{subfigure} 
	\caption{The first goal is to identify a specific accountability pattern in
	a given system model. }
	\label{fig:structure}
\end{figure}

Figure~\ref{fig:structure2}
shows the causal model for a fictitious definition of accountability
($\mathcal{D}$) and Figure~\ref{fig:structure1} shows a fictitious system model
($\mathcal{M}$). The question now is if this given model fulfills the given
definition of accountability. Unfortunately, we cannot simply compare the
graphs. Causal structures are more complex, and, as in this example, additional
intermediate nodes do not necessarily affect the equivalence of models (see
Appendix~\ref{sec:compare_causal_models} for details).  

\begin{figure}
	\centering
	\begin{subfigure}[t]{0.4\textwidth}
		\centering
		\includegraphics[height=.5cm]{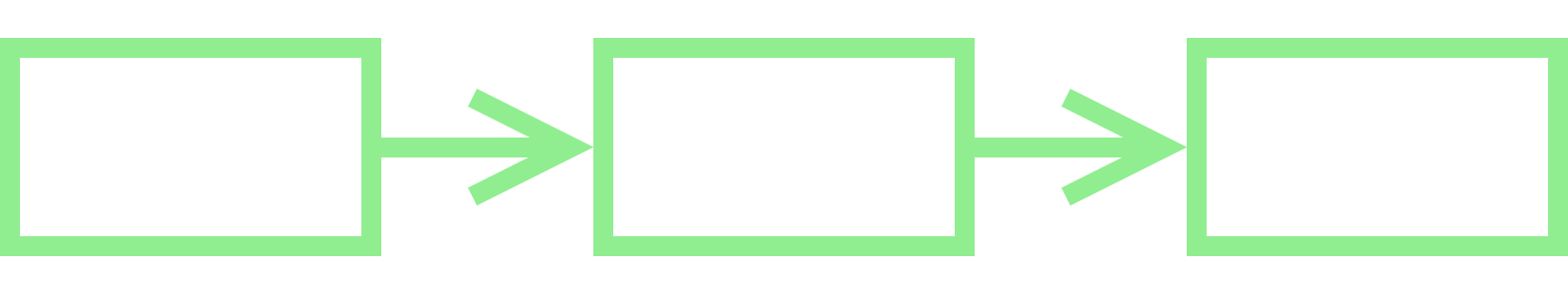}
		\caption{The causal model for this  definition of accountability
		requires a mediator.	}\label{fig:structure4}		
	\end{subfigure}
	\quad
	\begin{subfigure}[t]{0.4\textwidth}
		\centering
		\includegraphics[height=2cm]{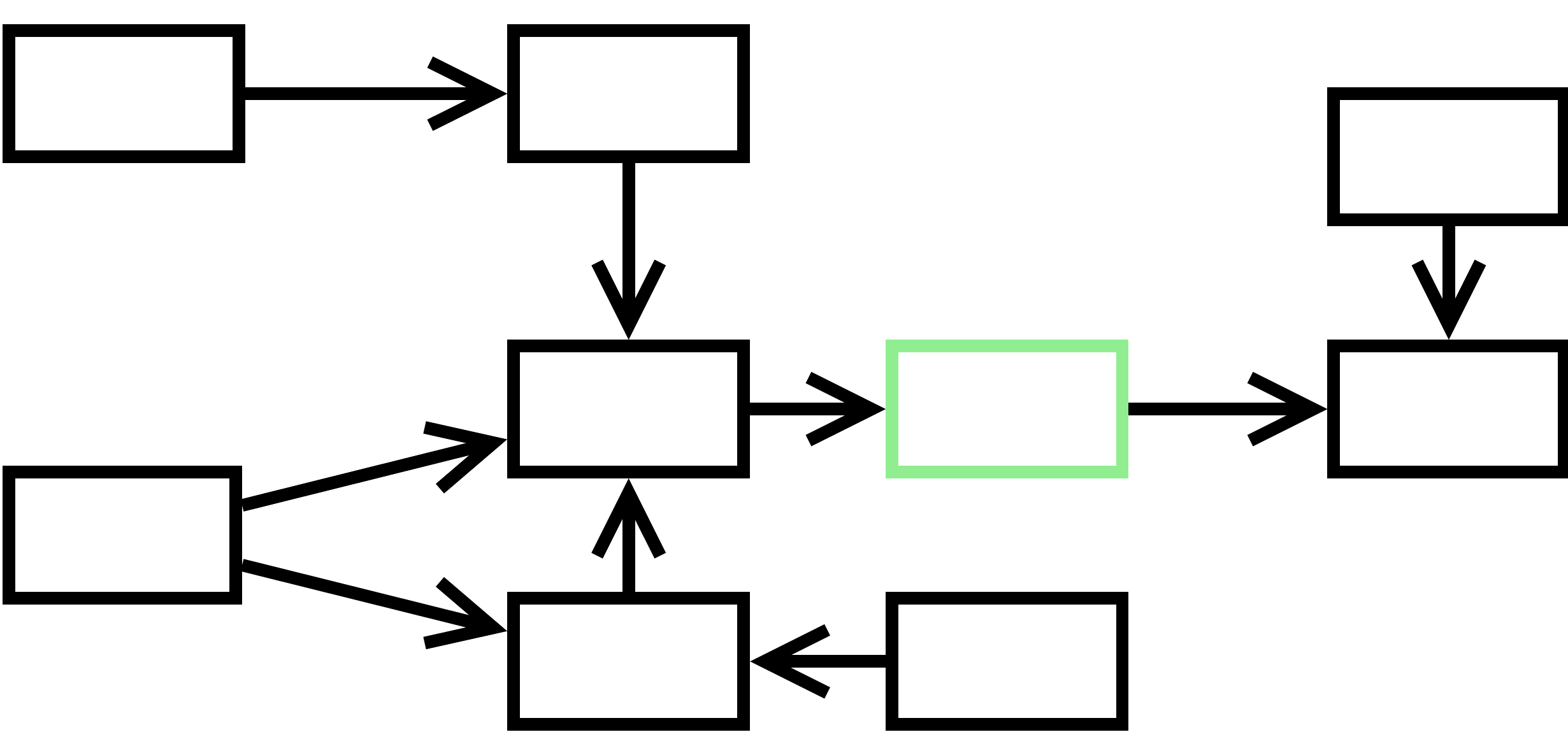}
		\caption{Adding the required node to the system model.}
		\label{fig:structure3}
	\end{subfigure} 
	\caption{Here, the problem is to change a system model to comply with a
	given definition of accountability. }
	\label{fig:structure_change}
\end{figure}

Figure~\ref{fig:structure_change} depicts another problem: How do we need to
alter a given system to comply with a given model of accountability? Here, the
fictitious accountability definition in  Figure~\ref{fig:structure4} requires a
mediator between the cause and an effect; a real-world example would be a person
confirming an order. Figure~\ref{fig:structure3} shows the model with such a
node added.

Looking at this another way, if we can show that the socio-technical system
$\mathcal{M}$ contains a causal structure $\mathcal{D}$, we can use the
accountability definition connected to $\mathcal{D}$ to make $\mathcal{M}$
accountable according to this definition. Conversely, if $\mathcal{M}$ does not
contain $\mathcal{D}$, it cannot be accountable according to this specific
definition.\footnote{But it might be accountable according to another
definition.}      In our experience, the following steps make for a
useful guideline to map accountability definitions onto causal models of
systems:\footnote{However, we have not proven these steps
to be the best approach. They are merely distilled from our experience; other,
possibly better approaches probably exist. }

\begin{enumerate}
\item Identify the event for which accountability is desired.  
\item Identify valid agents.
\item Choose the desired definition of accountability. 
\item For the given definition of accountability, check if the pattern is
fulfilled for the desired agents.
	\begin{enumerate}
		\item If the pattern is fulfilled, stop here.
		\item If not, change the model to fulfill the desired pattern.
	\end{enumerate}
\end{enumerate}

\section{Example}

We now use the 2018 deadly crash of an Uber car as an
example for the design decision in a system~\cite{elish2019moral,ntsb}. Here, we
will look at three different design choices for the control of the system and
reason about their accountability implications. In this accident, an autonomous
vehicle developed by Uber crashed into a pedestrian, Elaine Herzberg, crossing a
road and is regarded as the first accident in which a pedestrian was killed by
an autonomous vehicle. Ms.~Herzberg was pushing a bicycle while crossing a dimly
lit road and the software of the car repeatedly misclassified her, ultimately
hitting and killing her. The safety driver on board the vehicle was distracted
and did not brake in time. 

In the aftermath of the crash the accountability of the parties was hotly
contested. At first the police claimed it was the pedestrian's fault because at
the site of the accident, crossing the road was illegal. Next, the car's safety
driver was blamed because she did not pay attention to the road. The
manufacturer of the car's chassis, Volvo, was quick to distance itself from
any blame, arguing that its chassis had a collision avoidance system which
would have prevented the crash, but it was turned off by Uber to test their own
software. Velodyne, the manufacturer of the car's LiDAR, also pointed out that
their system was capable of detecting a pedestrian, but that their system does
not take the decision to brake. The search for reasons went as far as
criticizing Uber's development process, the testing process of having only one
driver in the car and even the car-friendly (and pedestrian-hostile) layout of
the road in Arizona or the point of autonomous cars in general. At the time of
writing, the safety driver is being indited with negligent
homicide~\cite{ubertrail}.
All these claims have in common that they ask counterfactual questions of a
causal model. Our goal is now to structure the causal model in such a way that
accountability can clearly be attributed. We  focus on a simplified model
that consists of three agents, namely Uber, who built the car, Volvo, who
contributed the chassis, and the safety driver, who was supervising the car. We
show how different SCMs give us different possible agents and how certain
structures of an SCM allow us to show that certain agents cannot be accountable
for a given outcome.

\subsection{Models of the System}
\begin{figure}
	\centering
	\begin{subfigure}[t]{0.45\textwidth}
		\centering
		\includegraphics[width=\linewidth]{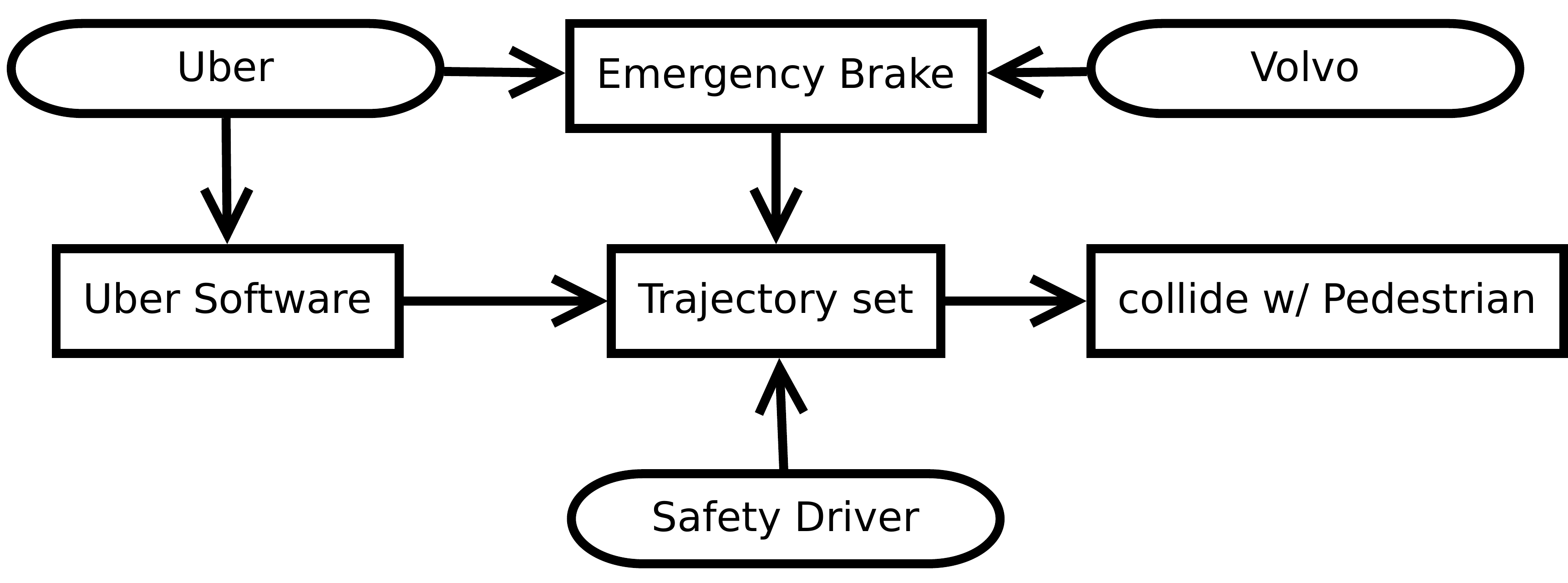}
		\caption{The human can take over. }\label{fig:uber1_2}		
	\end{subfigure}
	\quad
	\begin{subfigure}[t]{0.45\textwidth}
		\centering
		\includegraphics[width=\linewidth]{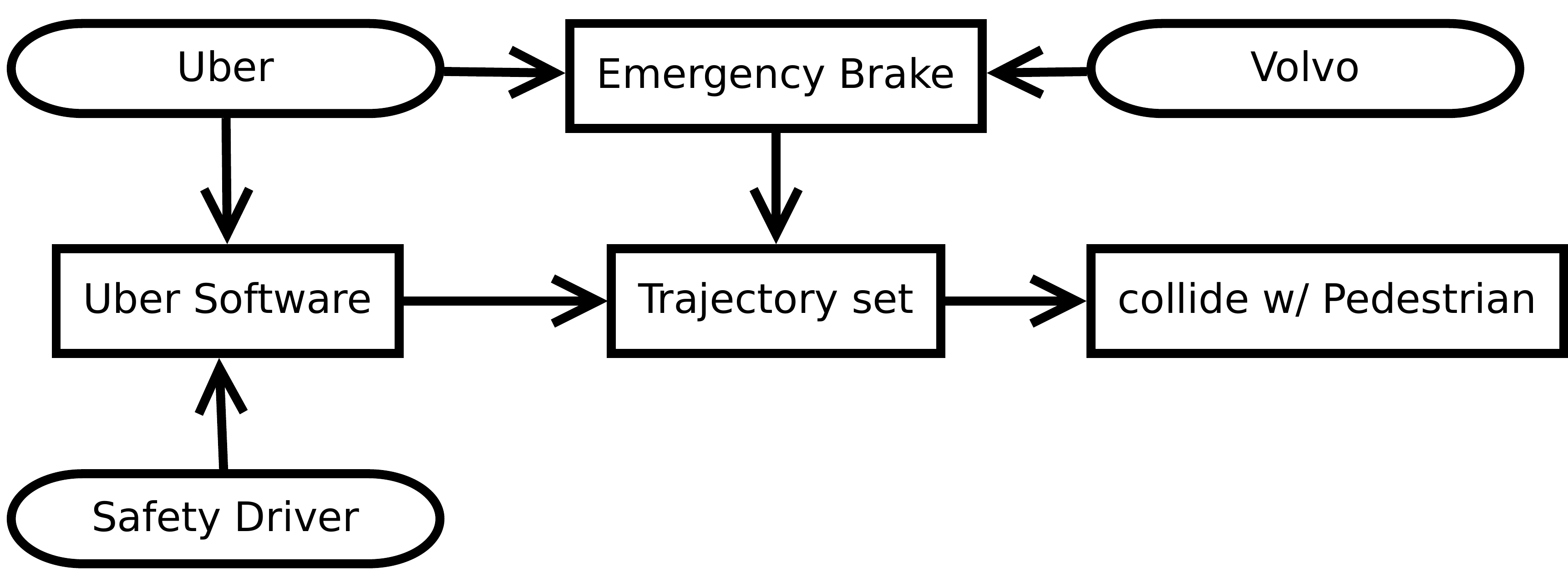}
		\caption{Human influence is moderated by the machine.
		}\label{fig:uber2_2}
	\end{subfigure} 
	\quad
	\begin{subfigure}[t]{0.45\textwidth}
		\centering
		\includegraphics[width=\linewidth]{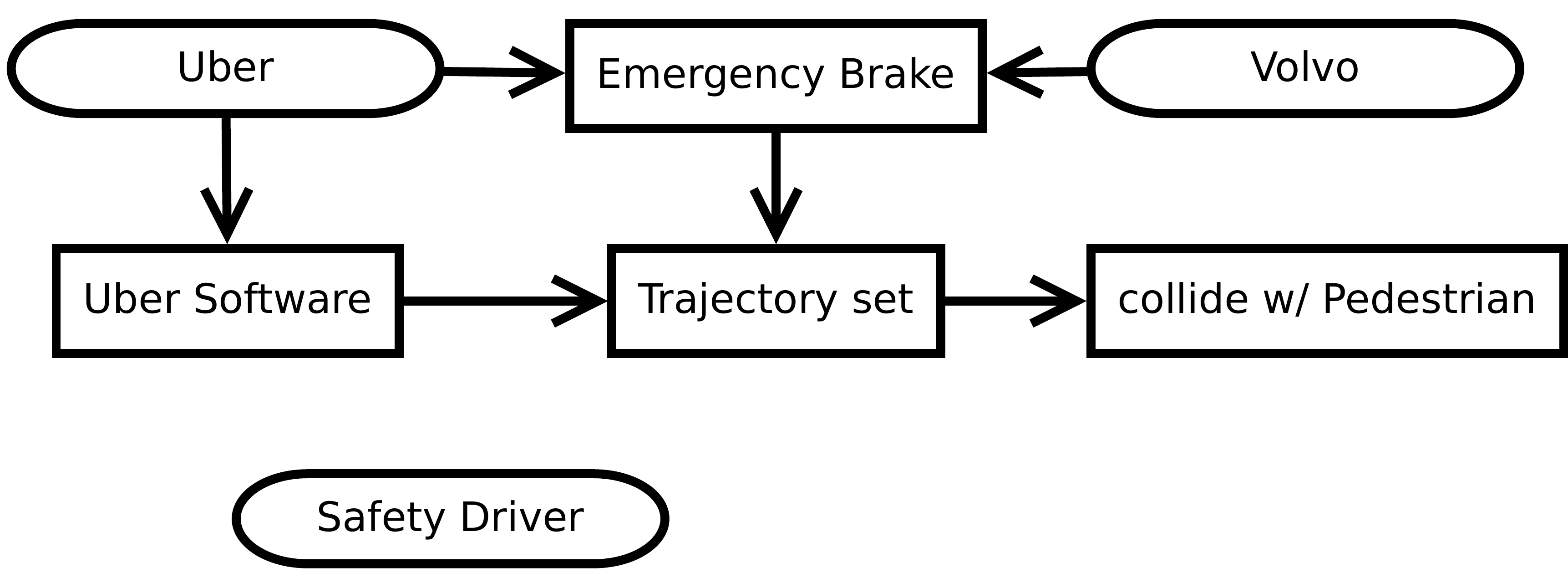}
		\caption{No human influence is possible. }\label{fig:uber3_2}
	\end{subfigure} 
	\caption{Three possible designs for a (semi-)autonomous car. While the
	SCMs show social entities, the system is not accountable as-is. }	
	\label{fig:two_uber2}
\end{figure}

Once we have the SCM $\mathcal{M}$ of the socio-technical system, we can then
use it to evaluate it for its accountability.  To illustrate this, we look at ways
to design an autonomous car (see Figure~\ref{fig:two_uber2}).  This example
illustrates three ways that  the control of such a system can be structured. In
Figure~\ref{fig:uber1_2}, the human can take over at any point in time,
Figure~\ref{fig:uber2_2} depicts a scenario where any input by the human can be
overridden by the machine, and Figure~\ref{fig:uber3_2} shows a setup where the
human cannot influence the car at all. In causal models, the lack of arrows
between two variables expresses the strong assumption that there is no causal
connection between these two variables. In these Figures we used rounded boxes
for possible agents (i.e., natural and legal persons) and rectangular boxes for
technical components. Here, we do not model any preemption. Temporal ordering
and the fact that one event might preempt another has a huge influence on the
model \cite{hitchcock2007prevention}.  

For simplicity, we assume this causal model to be binary.\footnote{This example,
is of course, highly simplified. In the real world, most causal models will not
be binary.} The meaning of the variables is as follows: 
\begin{enumerate}
\item \emph{collide w/ Pedestrian}, $P$, is true if a collision with a 
pedestrian occurs and false otherwise. 
\item \emph{Trajectory set}, $T$, is true if an evasive maneuver is conducted
and false  otherwise.
\item \emph{Safety Driver}, $D$, is true if the driver tries to change $T$ and 
otherwise false.
\item \emph{Uber Software}, $S$, is true if the car's software tries to change
$T$ and otherwise false.
\item \emph{Emergency Brake}, $E$, is true if the chassis tries to change $T$
and otherwise false.  
\item \emph{Volvo}, $V$, is true if $E$ is enabled and otherwise false. 
\item \emph{Uber}, $U$, is true if Uber influences $S$ or $E$. 

\end{enumerate}

Here it is important to note that we have a very lax approach to levels of
abstraction. Uber and Volvo are companies with unfathomable complexities, the
safety driver is a single individual, the software and the emergency brake are
complex technical systems, and the trajectory the outcome of multiple decisions.
However, SCMs can be abstracted quite well \cite{beckers2019abstracting} and so
this mixing of layers is easy to do formally. Still, it is important to bear in
mind that these variables will in reality be complex SCMs in their own
right. 

Formally, our model  $M = \mathcal{((U,V,R),F)}$ looks like this:
\begin{itemize}[leftmargin=*]
	\item[] $\mathcal{U} = \{\mathcal{U}_U,  \mathcal{U}_V, \mathcal{U}_D\}$ are
	the three exogenous variables. 
	\item[] $\mathcal{V} = \{P,T,D,S,E,V,U\}$ are the seven endogenous variables. 
	\item[] $\forall v \in \mathcal{U} \cup \mathcal{V} : \mathcal{R}(v) \to
	\{true, false\}$; since we assume a binary model, the range is
	$\{true,false\}$ for all variables.

\end{itemize}

Now we need to define $F_X$, so the structural equations for every
endogenous variable:
\begin{itemize}[leftmargin=*]

\item[] $U=\mathcal{U}_U$, $V=\mathcal{U}_V$, $D=\mathcal{U}_D$, meaning that
$U$, $V$, and $D$ are set by some exogenous variables.\footnote{This might seem
redundant, but the point is that only endogenous variables can be identified as
causes in a causal model. In real models, an endogenous variable will likely be
influenced by several exogenous variables. For example, the safety driver might
be influenced by blood alcohol level; other influences are the weather or the
road conditions.   }

\item[] $E = \left\{
\begin{array}{ll}
\lnot U &  U = true\\
V &  U = false \\
\end{array}
\right. $, so long as Uber is not disabling $E$, it is the value of $V$.

\item[] $S = U$ for Figure~\ref{fig:uber1_2} and Figure~\ref{fig:uber3_2}, says
that the software will follow whatever Uber had in mind. It is trickier for
Figure~\ref{fig:uber2_2}. Here we need to decide if $U$ or $D$ can override the
other and in what way. The simplest model would be a model in which the car will
break if either Uber or the safety driver wants to break. In this case  $S = U
\lor D$ would be the correct equation. If one could preempt the other, we would
need to change the model to contain such a preemption relation.

\item[] For Figure~\ref{fig:uber1_2} $T = S \lor E \lor D$ and
for the others $T = S \lor E $, if neither $S$, nor $E$, nor, as in
Figure~\ref{fig:uber1_2}, $D$ influence $T$, the car will hit the pedestrian.

\item[] $P = \lnot T $, if the car is on a collision course with the pedestrian
it will always hit her if the trajectory is not changed.

\end{itemize}

\subsection{Checking for Accountability}

We now have three distinct versions of $\mathcal{M}$. These models describe the
causal relations, but this alone is not sufficient for accountability. Our goal
now is to give a justification of why these options are accountable and decide
between the three.  First, we need to decide which event(s) we are concerned
about. In this example, we only care about potential collisions with
pedestrians. This means that we only care about causes that affect $P$. Next, we
need to identify valid agents.  Since accountability only has a meaning for
natural or legal persons, we can exclude any technical components in
$\mathcal{M}$, leaving us with three potential agents: Uber ($U$), Volvo ($V$),
and the safety driver ($D$). 

As the third step, we need to decide on a notion of
accountability~($\mathcal{D}$). Here we have two possibilities: Either
$\mathcal{D}$ is prescribed by some law or standard or we want to find a
$\mathcal{D}$ that is suitable for our socio-technical system. If we look at the
pattern given by the RACI definition, it is easy to see that $\mathcal{M}$
cannot be accountable in that sense. It is enough to look at the structure to
see that $\mathcal{M}$ simply does not have the necessary endogenous variables
to fulfill the RACI requirements. This is not surprising, as RACI is aimed at
organizations and a discussion is not something that translates well to
real-time systems like cars.  If we were under obligation to make that system
RACI-accountable, we can so show that this is impossible, given the current
$\mathcal{M}$. We would need to look for ways to extend $\mathcal{M}$ with the
required nodes. 

Making this system accountable according to Hall's definition would requires us
to convince the three agents that their behavior will potentially be evaluated
and that any misbehavior will be punished. In extremis, this would mean that if
we are convincing enough, we do not need to add any technical means of
accountability to our system. In practice, we would need to find a trade-off
between the cost of supervision and the compliance of the agents. For example,
it might make sense to collect all vehicle data to create a plausible scenario
of evaluation. To give an example, most people respect speed limits, despite the
fact that speed traps are quite rare. 

In many technical systems, the notions of Lindberg and Bovens will appeal to the
developers. In contrast to RACI, they do not require many additional nodes, and
in contrast to Hall, they both define a clear relationship between an agent (or
actor) and a principal (or forum). To apply their notion, we first need to
identify the basic structure they prescribe: $Agent \rightarrow Mediator
\rightarrow \textrm{\emph{Effect}}$.

In Figure~\ref{fig:uber1_2}, we can find the following causal chains leading to
the crash: $U \rightarrow S \rightarrow T \rightarrow P$, $U \rightarrow E
\rightarrow T \rightarrow P$, $V \rightarrow E \rightarrow T \rightarrow P$, and
$D \rightarrow T \rightarrow P$.  Here, the structure of the chain including the
safety driver is an exact match with this structure. For Uber and Volvo we have
two steps in the mediator and would need to show that they can be treated as a
single node. For Figure~\ref{fig:uber1_2} and Figure~\ref{fig:uber3_2} the
argument is straightforward because $S = U$, so these two nodes could be joined
into one. In Figure~\ref{fig:uber2_2}, we need to clarify if $U$ or $D$ has the
final say over the matter. As the model is given here, it would not be Lindberg
accountable because both $U$ and $D$ would be accountable for $T$. Similarly,
Volvo only has an influence on the trajectory if Uber is allowing them to do so.
So they fulfill this pattern if $U = false$.\footnote{Note that in
general showing that two causal models are equivalent is not trivial and cannot
be treated as a graph problem; see Appendix~\ref{sec:compare_causal_models}. }  

If we now take another look at $T$, we can see that its equation is as follows:
$T = S \lor E \lor D$. Again, we have the problem that we cannot disentangle the
effects from $U$, $E$, and $D$. However, given that $U$ will always disable $V$
(because $E = \lnot U$, if $U = true$), we can at least rule out $V$ as an
eligible agent. In Figure~\ref{fig:uber1_2} and Figure~\ref{fig:uber2_2}, it is
unclear if the safety driver or Uber are accountable, because both have a causal
effect on $T$. This agrees with the investigations in the aftermath of the
accident as there it was also unclear at first who was to be held accountable
for the deadly crash. Figure~\ref{fig:uber3_2} causality is much clearer because 
$D$ has no causal influence on $T$ at all.

Lastly, we will notice that none of the models has a principal. So we
need to determine who $A$ is accountable to. In the real world this role will be
filled with the authorities, so we would need to make sure that they have all
the evidence they need to understand the actions of $A$. So far, just judging
from the SCMs, we could either pick Lindberg's or Boven's definition.  Here, we
would pick Lindberg's because is has the notion that $P$ might inspect the
behavior of $A$ on demand. Boven's suggests regular reports, which seem
unnecessary for a car because accidents are rare. Boven's would be more suitable
if we were, for example, checking for violations of the speed limit. 

Given the three possible designs for the system, which would be the easiest to
make Lindberg accountable? Figure~\ref{fig:uber3_2}, where $D$ has no causal
influence and $V$ is inhibited by $U$, is attractive because only $U$ is left
as an agent. There would be no confusion about accountability. However,
Figure~\ref{fig:uber1_2} or Figure~\ref{fig:uber2_2} might be attractive in
practice because keeping the human-in-the-loop allows the technical system to
make wrong decisions without clear accountability of the manufacturer.

\subsection{Leveraging the Structure of $\mathcal{M}$}

Specific structures in causal models allow us to prove that one variable cannot
affect another in a specific way, even if there is a path between those two
variables. Two such structures are the Front and the Back-Door Criterion (see
Appendix~\ref{sec:analyzing}). Knowing that a specific variable has no causal
influence on another is invaluable for the design of a system, because this
means that we do not need to measure (or ``log'') it. This is helpful from an
engineering perspective, because it allows us to not store specific data and
thus save the cost for storage and the development of the data logging
functionality. It is also often desirable from a privacy perspective, because it
allows to justify not storing specific sensitive data. So if we can, for
example, show that skin color or religion have no causal influence, we can
justify not logging them, without compromising a system's accountability. To
return to our example in Figure~\ref{fig:uber3_2}, we might just be interested
in the effect of $S$ on $T$. The question now is, what other values do we need
to control for to calculate the effect of $S$ on $T$?\footnote{Here binary
models are not the best examples. It is easier to think of $S$ and $E$
contributing different amounts to a real valued function.} Employing the
Back-Door Criterion, we can see that we do have an open backdoor path, namely $S
\leftarrow U \rightarrow E \rightarrow T$, that will confound our estimate of
the effect of $S$ on $T$. To deconfound this reading, we could either control
for $U$, $E$, or both of them. This now allows us to justify not logging one of
these variables, provided we are only interested in the effect of $S$ on $T$.

\subsection{Actual Causality}

So far we used our models in a type causal manner, that is we looked at the
future, ensuring that the accountability for a specific event was clear and easy
to attribute. How would we  now use this model after an accident has actually
happened?  For this we can employ actual causality reasoning (see
Section~\ref{sec:actual}).\footnote{For ways to automatically check these
models, see \cite{ibrahim2019efficient,ibrahim2020actual,ibrahim2020from}. }
First, we need to ensure that we can set the context correctly. So in our
real-world system, we need sensors and logs that can tell us what has actually
happened.  Here we can leverage the fact that a causal model is determined by
its exogenous variables.\footnote{In this example we ignore any uncertainty. The
reasoning works similarly, but the results would be probability distributions.}
Looking at Figure~\ref{fig:two_uber2}, we have three exogenous variables:
$\mathcal{U}_U$ for Uber, $\mathcal{U}_V$ for Volvo, and $\mathcal{U}_D$ for the
safety driver.  All can be either $true$ or $false$.

If we measure  $\mathcal{U}_U = false,
\mathcal{U}_V = false, \mathcal{U}_D =
false$ and assume the model in Figure~\ref{fig:uber1_2}, will a crash occur? 
\begin{align*} 
E&= false \\
S&= U = false\\
T&= S \lor E \lor D = false \lor false \lor false = false\\
P&= \lnot T = true
\end{align*}
Here we can read this as \emph{Neither Uber nor the driver, nor Volvo tried to
change the trajectory, therefore the car crashed into the pedestrian}. However,
this sentence already includes the counterfactual assumption \emph{Had either Uber,
the driver, or Volvo done something else, the crash would not have happened}.
We can now formally check if this is correct. Since we assumed the model to be
binary, there is only one other thing the agents could have done, namely
whatever makes their ``measurement'' $true$.\footnote{This simplicity makes
binary models so popular for textbook examples. However, it is obvious that in
the real world defining ``something else'' will be tricky.} So, we can change
the value of $U$\footnote{Note that we just change the endogenous variable, not
the exogenous variable $\mathcal{U}_U$.} in the model and see if the result
changes. 

\begin{align*} 
E&= false \\
S&= U = true\\
T&= S \lor E \lor D = true \lor false \lor true = true\\
P&= \lnot T = false
\end{align*}
This setting can be read as \emph{Uber changed the trajectory and therefore the
car did not crash into the pedestrian}. So we can say that because there is a
counterfactual world in which $U$ could have prevented the crash, $U$ is a cause
for the crash to happen. 
If we now look at the causal model for Figure~\ref{fig:uber3_2}, we see that $T
= \lnot S \lor \lnot E$, so $D$ has no influence on $T$ (or any other variable
in the model). If $D = false$, we would get the same result as above. However,
if in the SCM above, we were to set $D = true$, $T$ would be $true$ and the
crash would be prevented. What would now happen in Figure~\ref{fig:uber3_2}?
\begin{align*} 
E&= false \\
S&= U = false\\
T&= S \lor E = false \lor false  = false\\
P&= \lnot T = true
\end{align*}
Despite the fact that we set $D = true$, so the driver tried to prevent the
accident in the counterfactual world, the accident still happens. This means
that $D$ is not a possible cause for the accident and, since causality is a
requirement for accountability, can also not be held accountable for the crash.  

\section{Conclusion}

Accountability is embedded deep into the fabric of society. Algorithmic systems
need to be designed in a way that conforms with these societal expectations.
This means that such important design decisions cannot be hidden deep within the
system. They need to be made explicit, communicated, and discussed. SCMs are
uniquely suitable for that because they allow us to formalize causality, the
necessary core of all definitions of accountability. 
In the current literature, SCMs are mainly used in scientific studies. The
models there are small and communicate assumptions about mechanisms in a study.
Developing SCMs for socio-technical systems is, despite some early work, still a
hard problem. Similarly, no clear-cut ways of identifying principals or
express definitions of accountability as SCMs exists. 

Despite these open problems, we are  convinced that SCMs offer the clarity that
is a requirement to make meaningful design decisions. While SCMs are not
sufficient to ensure accountability, a correct understanding of the underlying
causal mechanisms is necessary for any notion of accountability. Expressing this
as an SCM allows us to realize that we need certain structures in systems to
enable accountability. Without these structures,  a system cannot be
accountable. Once we have identified a specific structure, we can utilize
existing definitions of accountability and reuse the knowledge that comes with
them; we do not need to invent our own notions of accountability.   SCMs are a
powerful tool to analyze systems and, if they are not accountable,  provide a
well-reasoned argument why this is the case, and how the system should be
improved.

\begin{acks}
This work was supported by the Deutsche Forschungsgemeinschaft (DFG) under grant
no. PR1266/3-1, Design Paradigms for Societal-Scale Cyber-Physical Systems and
the Bavarian Research Institute for Digital Transformation (bidt).
\end{acks}
\bibliographystyle{ACM-Reference-Format}
\bibliography{bibliography}

\clearpage

\appendix

\section{Actual Causality}
\label{sec:hp}

The Halpern-Pearl (HP) definition\footnote{It is important to note that the HP
definition is just one possible way to define causality. As \cite[Ch.
2.2.2]{halpern2016actual} puts it so eloquently after introducing all the
details of the HP definition: ``At this point, ideally, I would prove a
theorem showing that some variant of the HP definition of actual causality is
the `right' definition of actual causality. But I know of no way to argue
convincingly that a definition is the `right' one; the best we can hope to do is
to show that it is useful.''}  uses the following formalization of a causal
model, based on Pearl's work on type causality \cite{halpern2015}:

\begin{definition}[Actual Causal Model] \label{def:cm}
	A  causal model $M$ is a tuple $M = \mathcal{((U,V,R),F)}$, where
	\begin{itemize}
	    \item $\mathcal{U}$ is a set of exogenous variables,
		\item $\mathcal{V}$ is a set of endogenous variables, 
		\item $\mathcal{R}:$ associates every variable with a nonempty set
		$\mathcal{R}(Y)$ of possible values $Y$,
		\item $\mathcal{F}$ associates with each variable $X \in
		\mathcal{V}$ a function that determines the value of $X$ (from the set 
		of possible values $\mathcal{R}(X)$) given the values of all other 
		variables
		$F_X : (\times_{U \in \mathcal{U}}\mathcal{R}(U)) \times (\times_{Y \in
			\mathcal{V}-\{X\}}\mathcal{R}(Y)) \to \mathcal{R}(X)$. 
	\end{itemize}

\end{definition}

A \textit{primitive event}, given $\mathcal{(U,V,R)}$, is a formula of the form
$X = x$ for $X \in \mathcal{V}$ and  $x \in \mathcal{R}(X)$. \textit{A causal
formula} is of the form $[Y_1 \leftarrow y_1, \dots, Y_k	\leftarrow
y_k]\varphi$, where $\varphi$ is a Boolean combination of primitive events.
$Y_1,\dots,Y_k$ (abbreviated $\overrightarrow{Y}$) are distinct variables in
$\mathcal{V}$, and $y_i \in \mathcal{R}(Y_i)$. Intuitively, this notation says
that  $\varphi$ would hold if $Y_i$ were set to $y_i$ for each $i$. $(M,
\overrightarrow{u}) \models X = x$ if the variable $X$ has value $x$ in the
unique solution  to the equations in $M$ in context $\overrightarrow{u}$ (i.e.,
the specific values of the variables). An intervention on a model is expressed
either by setting the values of $\overrightarrow{X}$ to $\overrightarrow{x}$,
written as $[X_1 \leftarrow x_1, .., X_k \leftarrow x_k]$, or by fixing the
values of $\overrightarrow{X}$ in the model, written as $M_{\overrightarrow{X}
\leftarrow \overrightarrow{x}}$. So, $(M, \overrightarrow{u}) \models
[\overrightarrow{Y} \leftarrow \overrightarrow{y}]\varphi$ is identical to
$(M_{\overrightarrow{Y} \leftarrow \overrightarrow{y}},\overrightarrow{u})
\models \varphi$.
 
 Using this definition of a causal model, an actual cause is defined as
 \cite{halpern2015}:
	
\begin{definition}[Actual Cause]\label{def:ac}
	$\overrightarrow{X} = \overrightarrow{x}$
	is an actual cause of $\varphi$ in $(M,\overrightarrow{u})$ if the following
	three conditions hold:\\ 
	\textbf{AC1.} \hspace{2mm} $(M,\overrightarrow{u}) \models
	(\overrightarrow{X} = \overrightarrow{x})$ and $(M,\overrightarrow{u})
	\models \varphi$.\\ 
	\textbf{AC2.} \hspace{2mm} There
	is a set $\overrightarrow{W}$ of variables in $\mathcal{V}$ and a setting
	$\overrightarrow{x}'$ of the variables in $\overrightarrow{X}$ such that if
	$(M,\overrightarrow{u}) \models \overrightarrow{W} = \overrightarrow{w}$,
	then $(M,\overrightarrow{u}) \models [\overrightarrow{X} \leftarrow
	\overrightarrow{x}', \overrightarrow{W} \leftarrow \overrightarrow{w}] \neg
	\varphi$.\\ 
	\textbf{AC3.} \hspace{2mm} $\overrightarrow{X}$ is minimal, i.e., no subset
	of $\overrightarrow{X}$ fulfills AC1 and AC2.
\end{definition}

Informally\footnote{For an in-depth discussion of all conditions and especially
AC2,\cite[Ch. 2.2.2]{halpern2016actual} is the most thorough resource.   } AC1
just says that a specific event $\overrightarrow{X} = \overrightarrow{x}$
actually
happened, otherwise it cannot be a cause. The minimality condition AC3 ensures
that only relevant events are part of a cause.  In Figure~\ref{fig:suzy_billy1},
for example, it would remove the detail that Alice is alive from the cause.  AC2
is the most complex condition and is thus traditionally explained last in any
text.  Here the idea is that we can show that $\phi$ depends on
$\overrightarrow{X}$ as long as we keep the variables in $\overrightarrow{W}$
fixed. This allows us to find that only the variables in $\overrightarrow{X}$
are affecting the outcome (and none of the variables in $\overrightarrow{W}$
do).

Coming back to the example depicted in Figure~\ref{fig:suzy_billy1}, we would
have three endogenous variables: $AT$ for \emph{Alice texted}, $BR$ for \emph{Bob runs
a red light}, and $AH$ for \emph{accident happens} as well the exogenous variables
$U_a$ and $U_b$ that set the value of the endogenous variables.  For simplicity,
all variables can be true or false. It is important to note that this example
assumes that there are no other factors that could influence the chain of
events. If there were, the model would be wrong and would need to be improved.
It is thus important that each causal model is discussed and ideally peer
reviewed. It is not enough to have a model: we also need a clear rational
for that model. Of course this will then cause second order questions of the
correctness of models, the bias of the modelers, and even who is allowed to
create the models.\footnote{See \cite{poechhacker2020} for some discussions of
this topic.} Here we simply assume that a causal model is a correct and detailed
enough representation of reality. 

\section{Special structures in SCMs}

\subsection{Chains, Forks, and Colliders}
\label{sec:chains}

\begin{figure}	
	\centering
	\begin{subfigure}[t]{1in}
		\centering
		\includegraphics[width=1in]{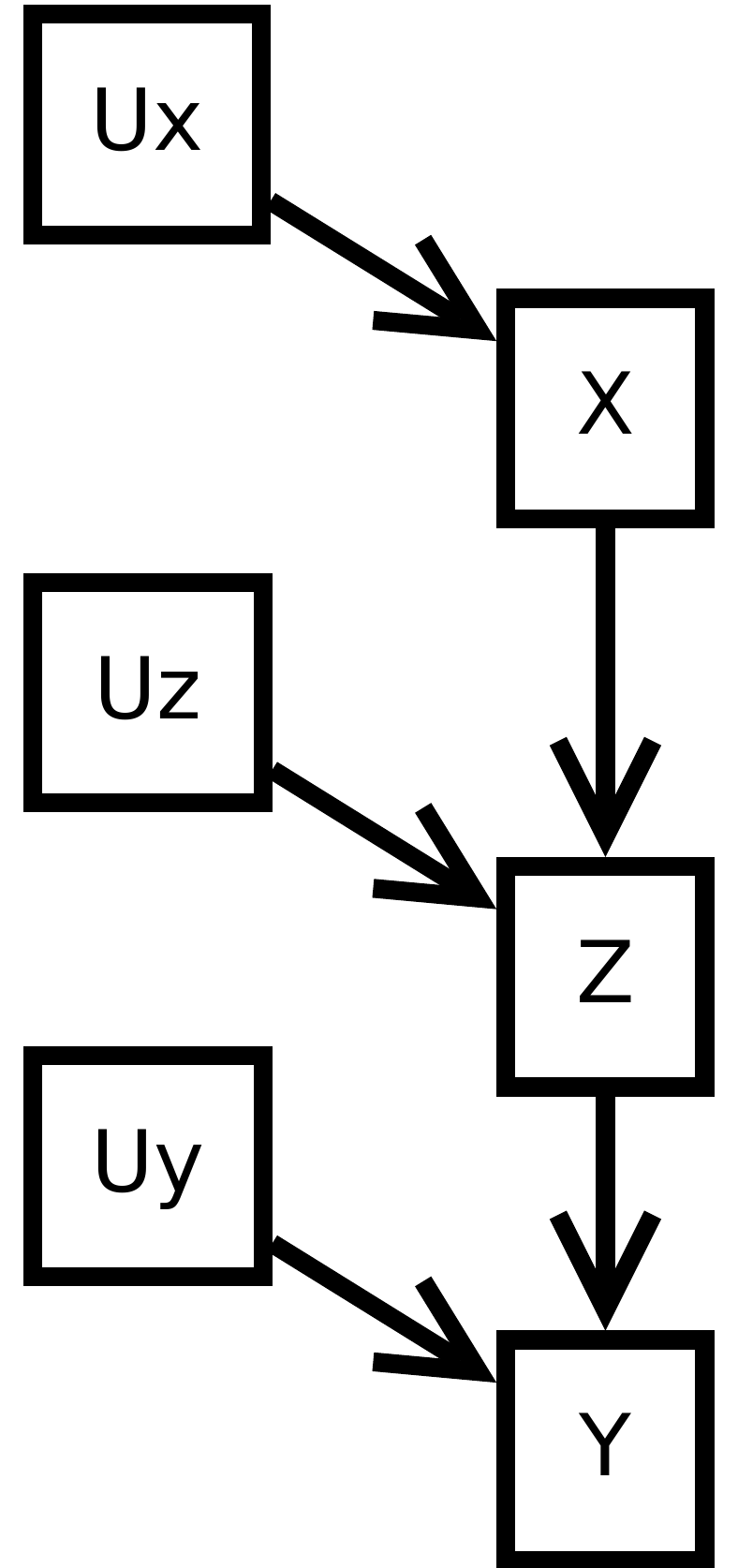}
		\caption{Chain.}\label{fig:chain}		
	\end{subfigure}
	\qquad
	\begin{subfigure}[t]{1in}
		\centering
		\includegraphics[width=1in]{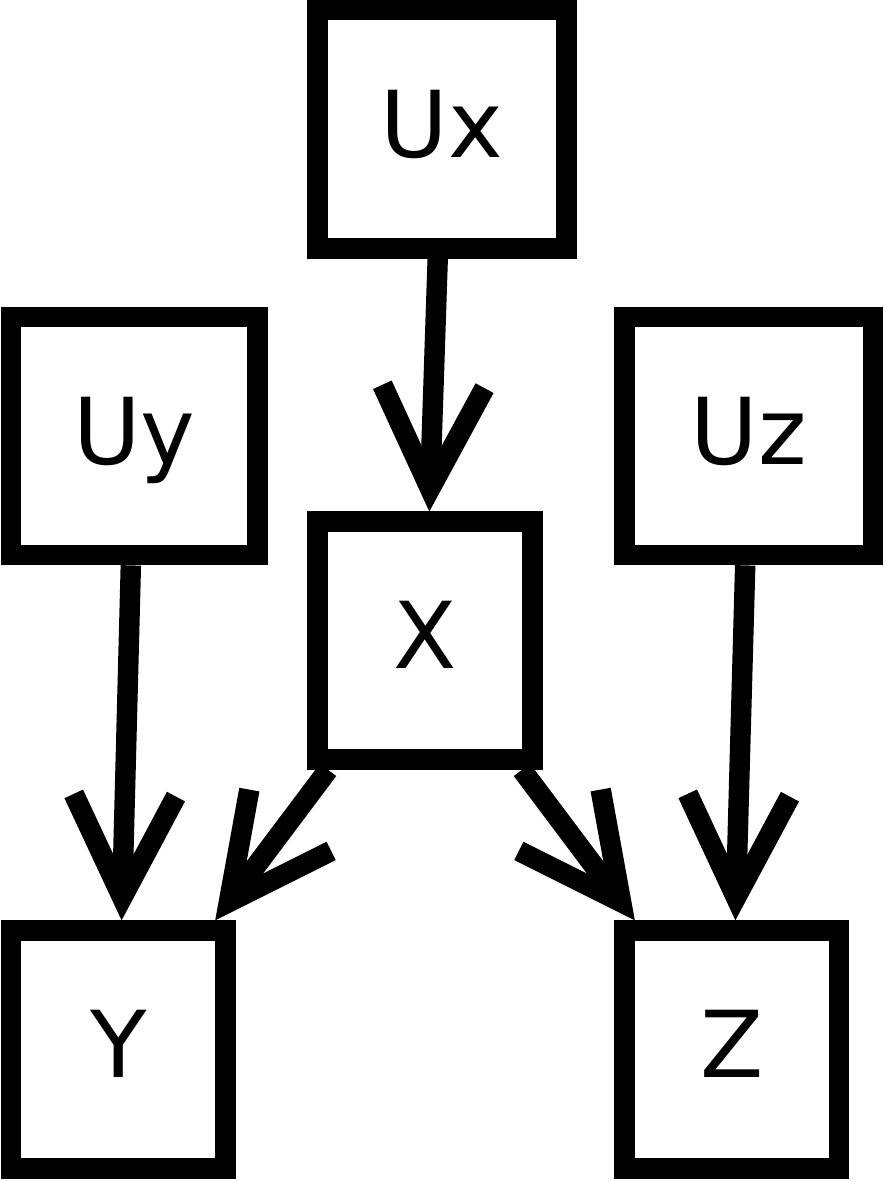}
		\caption{Fork.}\label{fig:fork}
	\end{subfigure}
	\qquad
	\begin{subfigure}[t]{1in}
		\centering
		\includegraphics[width=1in]{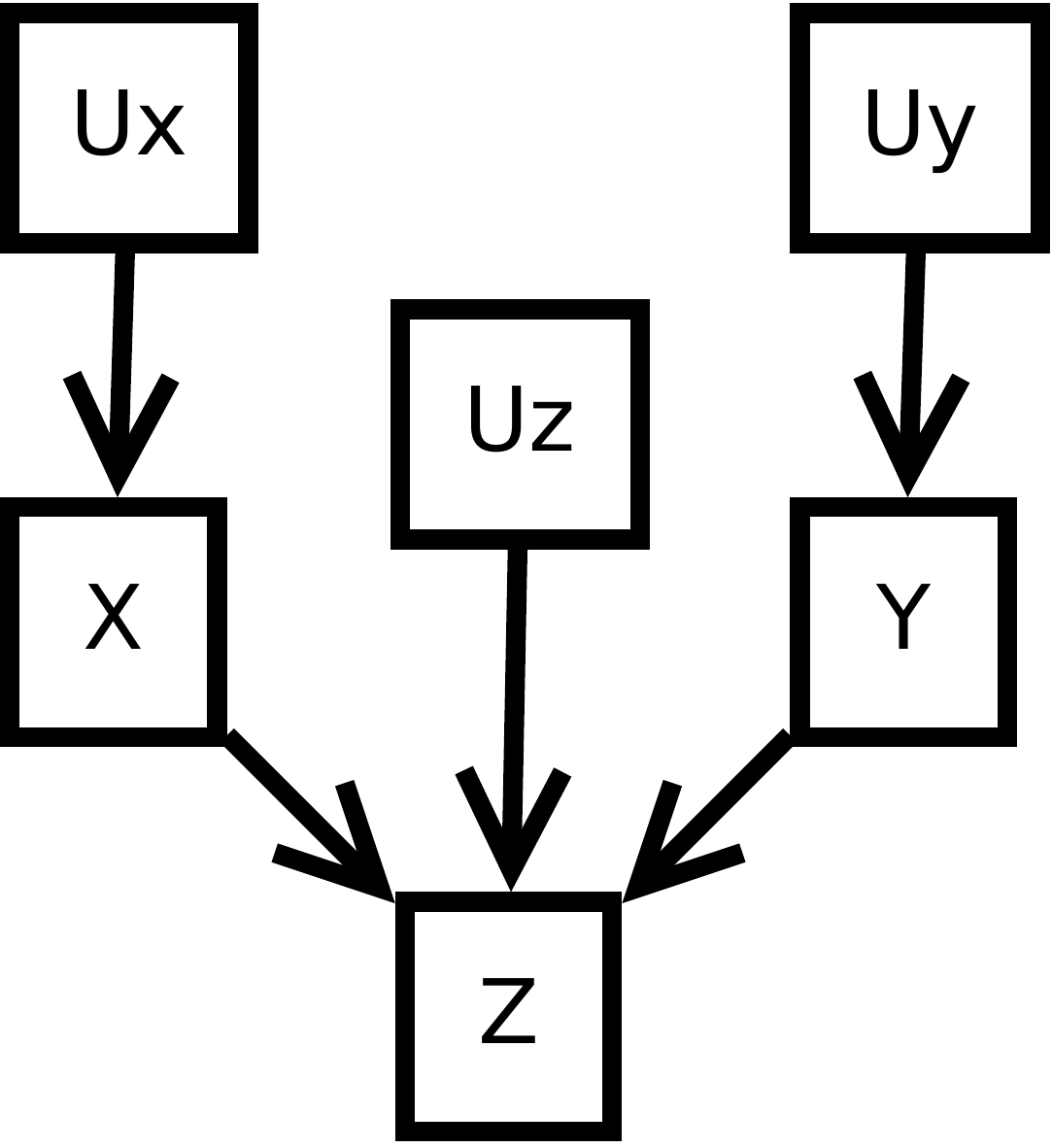}
		\caption{Collider.}\label{fig:collider}
	\end{subfigure}
	\caption{Three common structures in causal models. ${Ux}$, ${Uy}$, and
	${Uz}$ denote the exogenous variables. $X$, $Y$, and $Z$ are the endogenous
	variables. }\label{fig:basic_structures}
\end{figure}

In graphical causal models, some basic structures will arise repeatedly. Chains,
forks, and colliders exhibit very specific rules for causal
(in-)de\-pend\-en\-ce \cite[p.~35ff]{pearl2016causal}.
Figure~\ref{fig:basic_structures} depicts these three structures. In contrast to
the usual convention, we also model the exogenous variables ${Ux}$, ${Uy}$,
and ${Uz}$ here. In most causal models, only the endogenous variables, here  $X$,
$Y$, and $Z$ are modeled. Exogenous variables ``stand for any unknown or random
effect that may alter the relationship between endogenous variables''
\cite[p.~36]{pearl2016causal} and they are assumed to be independent of each
other. 

The important advantage of causal models is that in these structures several
(in-)de\-pend\-en\-cies hold, regardless of the function between those
variables. If we now look at Figure~\ref{fig:chain}, $Z$ is always dependent on
$X$. So, if we see the value of $X$ change, we will usually also see the value
of $Z$ change.  Next, $Y$ is \emph{likely} dependent on $X$. The reason for this
is that $Y$ depends on $Z$ for its value and $Z$ depends on $X$ for its value.
However, there are rare cases where changes in $X$ will not affect
$Y$.\footnote{See \cite[p.~38]{pearl2016causal} for an example.} And finally,
$X$ and $Y$ are independent, conditional on $Z$. The reason is that conditioning
on a variable means that we fix its value. If we had a dataset consisting of
three values, $X$, $Y$, and $Z$,  we would only look at the values of $X$
and $Y$, where $Z$ has a specific value. What happens here is that ${Uz}$
changes to keep $Z$ at this specific value. So whenever $X$ changes, ${Uz}$
would compensate for that change to keep $Z$ constant and as $Y$ only depends on
$Z$ and not on ${Uz}$, its value is independent of the changes in $X$. 

Figure~\ref{fig:fork} depicts a fork. Here, $Y$ and $Z$ both depend on $X$ and
this also means that $Y$ and $Z$ are \emph{likely} dependent. The reason is
that since both depend on $X$, a change in one will inform us that the other
will also likely change. However, there are also cases where this is not the
case. Finally, $Y$ and $Z$ are independent, conditional on $X$. Similar to the
chain above, once we hold $X$ constant, a change in either $Y$ or $Z$  no
longer indicates a change in the other. In this structure, $X$ is called a
\emph{common cause}. 

The third basic structure, a collider, is depicted in Figure~\ref{fig:collider}.
Here, we can see that $Z$ is dependent on $X$ and $Y$ and $X$ and $Y$ are
independent, because they are not in a parent-child relationship and their
exogenous variable $Ux$ and $Uy$ are assumed to be independent. The most
surprising property of a collide is that $X$ and $Y$ become dependent,
conditional on $Z$. While it might be surprising that two independent variables
can suddenly become dependent, it can be illustrated with a simple example
\cite[p.~41]{pearl2016causal}: If, we assume $X + Y = Z$ and $X$ and $Y$ are
independent, knowing that $X = 3$ does not tell you anything about $Y$. However,
the moment you know that $Z = 10$, knowing that $X = 3$ lets you deduce that $Y
= 7$.

This concept of  (in-)dependence can now be generalized for all graphs with the
concept of d-separation. This means that two variables are independent if every
path between them is \emph{blocked}.  Formally \cite[p.~46]{pearl2016causal},

\begin{definition}[d-seperation]
\label{def:dsep}

A path $p$ is blocked by a set of nodes $Z$ if and only if
\begin{enumerate}
\item $p$ contains a chain of nodes $A \rightarrow B \rightarrow C$, or a fork 
$A \leftarrow B \rightarrow C$, such that the middle node $B$ is in $Z$ (i.e.,
$B$ is conditioned on), or
\item $p$ contains a collider  $A \rightarrow B \leftarrow C$ such that the
collosion node $B$ is not in $Z$, and no descendant of $B$ is in $Z$. 
\end{enumerate}

If $Z$ blocks every path between two nodes $X$ and $Y$, then $X$ and $Y$ are
d-separated, conditional on $Z$, and thus are independent conditional on $Z$.  

\end{definition}

\subsection{Analyzing Causal Models}
\label{sec:analyzing}

\cite[p.~157]{pearl2018book} distills these properties of causal models into
four rules: 

\begin{itshape}
\begin{enumerate}
\item In a chain junction, $A \rightarrow B \rightarrow C$, controlling for $B$
prevents information about $A$ from getting to $C$ or vice versa.

\item Likewise, in a fork or confounding junction, $A \leftarrow B \rightarrow
C$, controlling for $B$ prevents information about $A$ from getting to $C$ and
vice versa. 

\item In a collider, $A \rightarrow B \leftarrow C$, exactly the opposite rules
hold. The variables $A$ and $C$ start out independent, so that information about
$A$ tells you nothing about $C$. But if you control for $B$, then information
starts flowing through the ``pipe'', due to the explain-away effect. 

\item Controlling for descendants (or proxies) of a variable is like
``partially'' controlling for the variable itself. Controlling for a descendant
of a mediator partly closes the pipe; controlling for a descendant of a collider
partly opens the pipe.

\end{enumerate}
\end{itshape}

Even if we have a long causal path, it is enough that one junction blocks the
information flow. To deconfound two variables, we need to block any noncausal
path while not blocking any causal path. This leads to two prominent criteria to
identify causal independence: The Back-Door and the Front-Door
Criterion.\footnote{All of the following examples are based on examples given by
\cite{pearl2018book}; \cite{tikka2018identifying} provide a tool to automate the
analysis.}

\subsubsection{The Back-Door Criterion}

\begin{definition}[The Back-Door Criterion]
Given an ordered pair of variables
$(X,Y)$ in a directed acyclic graph $G$, a set of variables $Z$ satisfies the
Back-Door Criterion relative to $(X,Y)$ if no node in $Z$ is a descendant of
$X$, and $Z$ blocks every path between $X$ and $Y$ that contains an arrow into
$X$.  \end{definition}

Intuitively, the Back-Door Criterion \cite[p.~61]{pearl2016causal}
ensures that (1) all spurious paths between
$X$ and $Y$ are blocked, (2) all directed paths from $X$ to $Y$ are not
perturbed, and (3) no new spurious paths are
added.\footnote{\cite{greenland1999causal} give a detailed and in-depth
explanation of the Back-Door Criterion and why it works on graphical models. }

\subsection{The Front-Door Criterion}
\label{sec:front-door}

\begin{definition}[The Front-Door Criterion]
A set of variables $Z$ is said to satisfy the Front-Door Criterion relative to
an ordered pair of variables $(X,Y)$ if
\begin{enumerate}
\item $Z$ intercepts all directed paths from $X$ to $Y$.
\item There is no unblocked path from $X$ to $Z$.
\item All back-door paths from $Z$ to $Y$ are blocked by $X$.
\end{enumerate}

\end{definition}

Intuitively, the Front-Door Criterion \cite[p.~69]{pearl2016causal}
relies on the fact that one can identify
the effect of $X$ on $Z$ and the effect of $Z$ on $Y$ separately. This works
because $Z$, so the mechanism (or mediator) by which $X$ affects $Y$, is not
affected by any unobserved confounders.\footnote{The Front-Door Criterion will
also work if $Z$ is only weakly affected by a confounder. The results will,
however, get more imprecise the bigger $Z$ is influenced. }  Having
identified the separate effect, we can then calculate the effect from $X$ on
$Y$.\footnote{See \cite{bellemare2019paper} for an in-depth discussion of the
application of the front-door criterion.}

\subsection{Comparing an SCM to an Accountability Definition}
\label{sec:compare_causal_models}

In our approach, we  end up with two causal models. $\mathcal{M} =
\mathcal{((U,V,R),F)}$ is an SCM of the system that should be accountable and
$\mathcal{D} = \mathcal{((U',V',R'),F')}$ is an SCM of an accountability
definition. Here it is very important to notice that in general the signatures
of the two models will be different, i.e.  $\mathcal{(U,V,R)} \neq
\mathcal{(U',V',R')}$. The reason for this is that the model of a system will
usually contain more nodes than those of a definition.  This is in so far a
problem because much of the literature on the equivalence of causal models is
concerned with so called Markov Equivalence Classes\footnote{See for example
\cite{verma1992algorithm,andersson1997characterization,radhakrishnan2016counting}
and \cite{jaber2019causal} for recent advances.}, which requires the signatures
of the models to be identical. This, however, is not useful in comparing
accountability models to system models because their signature will always be
different. $\mathcal{D}$ is supposed to be a small model focused on the
important aspects of accountability, while    $\mathcal{M}$ is supposed to
represent a whole system.  

\begin{figure}
	\centering
	\begin{subfigure}[t]{0.2\textwidth}
		\centering
		\includegraphics[height=0.5cm]{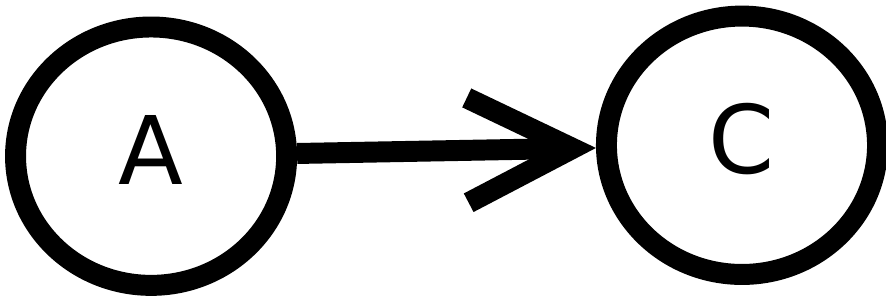}
		\caption{$\mathcal{F} = \{C=A\}$   }
		\label{fig:equi1}		
	\end{subfigure}
	\quad
	\begin{subfigure}[t]{0.2\textwidth}
		\centering
		\includegraphics[height=0.5cm]{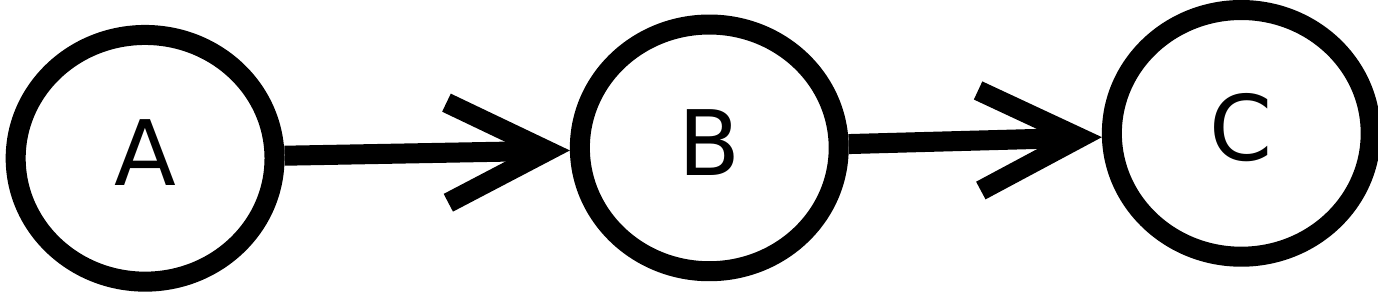}
		\caption{ $\mathcal{F} = \{C=B,B=A\}$}
		\label{fig:equi2}		
	\end{subfigure} 
	\caption{Two models that are equivalent with regard to $A$ and $C$.   }
	\label{fig:equi_two}
\end{figure}

\begin{figure}
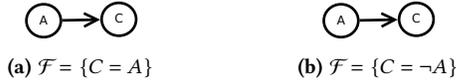

	\centering
	\begin{subfigure}[t]{0.2\textwidth}
		\centering
		\includegraphics[height=0.5cm]{equi1.pdf}
		\caption{$\mathcal{F} = \{C=A\}$   }
		\label{fig:equi3}		
	\end{subfigure}
	\quad
	\begin{subfigure}[t]{0.2\textwidth}
		\centering
		\includegraphics[height=0.5cm]{equi1.pdf}
		\caption{$\mathcal{F} = \{C=\lnot A\}$   }
		\label{fig:equi4}		
	\end{subfigure} 
	\caption{Despite their similar structure, these two models are not
	equivalent.   }
	\label{fig:equi_similar}
\end{figure}

Figure~\ref{fig:equi_two} depicts two causal models. In Figure~\ref{fig:equi1},
$A$ causes $C$ directly, whereas in Figure~\ref{fig:equi2} $A$ causes $C$ via a
mediator $B$. The question now is if these two models are equivalent and if so
what this means. If we just look at the graph structure, these two models are
different. One has three endogenous variables and the other only two. However,
if we look at structural equations of this model, we can see that $B$ does not
affect the influence of $A$ on $C$. So any useful notion of equivalence would
need to find that these two models are equivalent. If it would not, this notion
of equivalence would be next to useless because one can always include
additional intermediate variables in any causal model.
Figure~\ref{fig:equi_similar} depicts two models that have a similar
structure, but are functionally their complete opposite. Any notion of
equivalence should find these two models distinct.

\end{document}